\documentclass[preprint,aps,prc,showpacs]{revtex4}

\usepackage{graphicx}% Include figure files
\usepackage{bm}% bold math

\begin{document}

\title{ Three-body collisions in Boltzmann-Uehling-Uhlenbeck
        theory% 
        \footnote{Work supported by GSI Darmstadt} }
\author{A.B. Larionov$^{1,2,3}$, O. Buss$^1$, K. Gallmeister$^1$ and U. Mosel$^1$}

\affiliation{$^1$Institut f\"ur Theoretische Physik, Universit\"at Giessen,
             D-35392 Giessen, Germany\\
             $^2$Frankfurt Institute for Advanced Studies, J.W. Goethe-Universit\"at,
             D-60438 Frankfurt am Main, Germany\\ 
             $^3$Russian Research Center Kurchatov Institute, 
             123182 Moscow, Russia}

\date{\today}

\begin{abstract}
Aiming at a microscopic description of heavy ion collisions in the beam energy
region of about 10 A GeV, we extend the Giessen Boltzmann-Uehling-Uhlenbeck
(GiBUU) transport model by including a relativistic mean field, in-medium baryon-baryon
cross sections and three-body collisions. The model is then compared with
experimental data for central Au+Au collisions at 2-10 A GeV and central Pb+Pb
collisions at 30 and 40 A GeV on the proton rapidity spectra, the midrapidity yields
of $\pi^+$, $K^\pm$ and $(\Lambda+\Sigma^0)$, and  the transverse mass spectra
of $\pi^\pm$ and $K^\pm$. The three-body collisions increase the inverse slope
parameters of the hadron $m_\perp$-spectra to a good agreement with the data.
\end{abstract}


\maketitle

\section{ Introduction }

The Boltzmann-Uehling-Uhlenbeck (BUU) transport theory, based on binary collisions  
and on the propagation of particles in a selfconsistent mean field, is
a very useful tool in understanding the heavy ion collisions (HIC) in the
energy region from the Fermi energies ($E_{\rm lab} \sim 30$ A MeV)
up to the relativistic energies ($E_{\rm lab} \sim 2$ A GeV) (c.f. Refs.
\cite{BDG88,CMMN90,Teis97,CB99,EBM99,MEPhD,GiBUU} for the description of the model in various numerical
realizations and Refs. \cite{LCGM00,LM03,THW05,LM05} for the results below 2 A GeV).

There are, however, systematic deviations from experimental data on 
pion and kaon production at $E_{\rm lab} > 2$ A GeV \cite{Weber03,Brat04,WLM05}:
the pion multiplicity is systematically overpredicted, while the slopes of
the $K^+$ transverse mass spectra are too steep in the transport calculations.
It has been advocated in \cite{Brat04}, that the too soft kaon $m_\perp$-spectra are
caused by missing the formation of a nonhadronic phase, which should create an
additional pressure accelerating the kaons. On the other hand, the three-fluid
hydrodynamical calculations \cite{IR06} have apparently been quite successfull
in reproducing the $m_\perp$-spectra of hadrons at AGS to SPS energies by
using a hadronic equation of state. This points to the idea of not enough 
thermalization produced by the microscopic transport models rather than to the 
importance of the nonhadronic degrees of freedom.   

The many-body collisions --- usually missed in the current transport calculations at high energies
---  could serve as an additional source of the thermalization.
The role of many-body collisions grows with baryon density, which
reaches values of order of 1-2 fm$^{-3}$ at the energy region of 10-20 A GeV  
\cite{Randrup06}. These are typical energies of the future Compressed Baryonic Matter 
experiment at the Facility for Antiproton and Ion Research in Darmstadt \cite{FAIR}.

A simple estimate of the gas parameter \cite{LP} at the baryon density $\rho_B=10\rho_0$, which 
is the maximum density reached in a central Au+Au collision at 20 A GeV with 
$\rho_0=0.17$ fm$^{-3}$ being the normal nuclear matter density, is
\begin{equation}
    \gamma_{\rm gas} = (\sigma/\pi)^{3/2} \rho_B \simeq 2~,            \label{gammaGas}
\end{equation}
where $\sigma \simeq 40$ mb is the asymptotic high-energy value of the total pp cross section
in vacuum. In thermally equilibrated nuclear matter, one can neglect the Lorentz contraction
of the interaction volume $\frac{4}{3}\pi(\sigma/\pi)^{3/2}$ (c.f. Eqs.(\ref{V12}),(\ref{V12_aver}) 
and Fig.~\ref{fig:freq} below).
Since $\gamma_{\rm gas} > 1$, one concludes, that the applicability condition
of the binary collision approximation is violated (c.f. Ref. \cite{Mrow85}). 
As we have recently shown \cite{LGM06}, at the maximum compression stage of the central 
Au+Au collision at 20 A GeV, $N$-body collisions with $N \geq 6$ should dominate.

One way to describe this complex physical situation could be, indeed, an introduction of the 
new degrees of freedom, e.g. like the formation of a quark-gluon plasma. This is quite 
a challenging problem for the microscopic transport theories, which is, however, beyond
the scope of our present study. Another way is to simulate the many-body collisions within a 
theory containing only the hadronic degrees of freedom. At beam energies below 1-2 A GeV,
transport models taking into account $N$-body collisions (with $N \geq 3$) have been constructed 
by several authors (c.f. Refs. \cite{Kodama82,BRV92,BM93,BGM94,WH95}). 
The difficulty, which appears at high baryon densities reached in the 10 A GeV region is that the 
gas parameter is not small, and, therefore, such a theory can not be formulated with the vacuum cross 
sections as a decomposition of the collision integral in a series of powers of $\gamma_{\rm gas}$
(c.f. Ref. \cite{LP}). In other words, the vacuum cross sections should be screened at high densities 
by particles surrounding the colliding pair \cite{Dani00}. The screening effect appears also as
a consequence of the Dirac-Brueckner calculations (c.f. Refs. \cite{tHM87,Fuchs01}), where
the nuclear mean field and the in-medium reduced cross sections are derived from the same 
fundamental vacuum interaction. Using in-medium reduced cross sections would reduce
the relative contribution of the many-body collisions, since the gas parameter will be smaller.
This would give a more solid ground to the kinetic theory, which is based on the small
parameter $\gamma_{\rm gas}$. The Fermi liquid theory \cite{Landau,Pomeranchuk} gives 
a similar picture. There, the liquid of the real particles is equivalently replaced by the gas
of quasiparticles. 

In the present work, we develop a transport model which contains a Walecka-type
baryonic mean field and the in-medium reduced baryon-baryon cross sections.
The model is then extended by including three-body collisions. The mean field and 
the in-medium reduced cross sections lead to less thermalization, while the three-body 
collisions counterbalance this effect.
We will show that the three-body collisions raise the inverse slope parameters
of the hadron spectra in the central HIC at beam energies of 2-40 A GeV.
In particular, the measured $K^+$ transverse mass spectra are well described
by calculations with the three-body collisions.    
      
The structure of the paper is as follows. In Sect. II we briefly describe the GiBUU 
model \cite{GiBUU} concentrating  on its new ingredients: the relativistic mean field, 
the in-medium baryon-baryon cross sections and the three-body collisions. 
Sect. III contains numerical results. In Sect. IV we summarize and discuss our results.

\section{ GiBUU model }

Our calculations are based on the GiBUU model in a new version of Ref. \cite{GiBUU},
written using FORTRAN 2003 in an object-oriented way.
The model describes a nucleus-nucleus collision explicitly in time as a sequence of the baryon-baryon,
meson-baryon and meson-meson collisions, baryon and meson resonance excitations
and decays. Between the two-body collisions the particles propagate in a
selfconsistent mean field.  The baryon-baryon collisions at $\sqrt{s} \leq 2.6$ GeV
are treated within the resonance model, while at $\sqrt{s} > 2.6$ GeV the 
FRITIOF model is applied. For the meson-baryon collisions, the FRITIOF
model is used at $\sqrt{s} > 2.2$ GeV. We applied the energy-dependent strangeness
suppression factor 
\begin{equation}
              \gamma_s \equiv \frac{P(s)}{P(u)} =
\cases{
0.4 & for $\sqrt{s}\leq 5$ GeV \cr
0.433-\frac{1}{150}\sqrt{s}\,\hbox{[GeV]}^{-1} & 
for 5 GeV $<\sqrt{s}<$ 20 GeV \cr
0.3 & for $\sqrt{s}\geq 20$ GeV \cr
}
                                                             \label{strsuppr}
\end{equation}
from Ref. \cite{Geiss} instead of the default FRITIOF value of $\gamma_s = 0.3$.
In the meson-baryon collision case, the FRITIOF mechanism of the double-string excitation
and decay has been improved by adding the $q \bar q$ annihilation channel as described in
Ref. \cite{WLM05}. 

The particles produced in string decays are not allowed to interact with their default
cross sections up to some proper time interval, called formation time. We use a value
of the formation time $\tau_f=0.8$ fm/c for all baryons and mesons. If one of the
colliding particles is still in the formation interval, we call it a prehadron. The total cross section
of the prehadron interaction with another particles is scaled according to the constituent
quark model (c.f. Ref. \cite{Falter04}): If the prehadron interacts with a hadron, the
total cross section is multiplied by the factor of $N_{\rm leading}/3$ ($N_{\rm leading}/2$)
for the pre-baryon (pre-meson), where $N_{\rm leading}$ is the number of leading (anti)quarks
in the prehadron, i.e. the number of (anti)quarks which were existing in the parent colliding 
particles for the given prehadron. If both colliding particles are prehadrons,
the total cross section of their interaction is multiplied by the product of the two corresponding
factors. 

The rescaling factor of the prehadron cross sections is quite important for the description of
the produced particle multiplicities. In previous GiBUU calculations of HIC at 2-40 A GeV
\cite{WLM05}, this factor had been chosen to be 1/3 for the prehadrons containing at least one
leading quark and zero otherwise. This explains the higher particle abundancies produced in the
cascade mode in the present paper as compared to the results of Ref. \cite{WLM05}.
Another distinction to the calculations of Ref. \cite{WLM05} is that in the present work
we miss the low energy ($\sqrt{s} \leq 2.6$ GeV) baryon-baryon  channel of kaon production
$BB \to BYK$. However, this channel was found to be negligible for the beam energies above 
4 A GeV \cite{WLM05}. 

In this work, we improve the previous GiBUU calculations \cite{WLM05} by implementing
a relativistic mean field (RMF), in-medium cross sections and three-body collisions in the model. 
The in-medium cross sections for the high energy baryon-baryon
collisions computed according to the RMF model have already been introduced in \cite{WLM05}.
However, the RMF has not been used for the particle propagation in \cite{WLM05}. 
Below, we describe in-detail these new ingredients of our model.

\subsection{Relativistic mean field}

In distinction to the earlier GiBUU calculations \cite{EBM99,MEPhD,LCGM00,LM03,LM05,Falter04},
where the nonrelativistic momentum-dependent potential of Ref. \cite{Welke88} was used, 
in the present work we employ the relativistic treatment of the baryonic mean field.
In spite of the Lorentz invariant implementation of the potential from \cite{Welke88}, when it is 
calculated in the local rest frame of nuclear matter, the RMF is better suited for description of the 
high baryon densities reached in a central heavy ion collision at about 10 A GeV beam energy. 
Here, a consistent nuclear equation of state at high densities is of the primary importance. RMF models 
provide a quite good description of both the collective flow in HIC and of the high density interior of  
neutron stars \cite{Blaettel93,Klaehn06}. They are also successfully applied in nuclear structure 
calculations \cite{LKR97}. The one weakness of the RMF models, the too repulsive nucleon-nucleus
interaction at high momenta (c.f. \cite{Blaettel93} and refs. therein) , is of minor relevance for 
the present results for particle production from high-density equilibrated nuclear matter.

We use the relativistic mean field Lagrangian density in the form given in Ref. \cite{LKR97}:
\begin{eqnarray}
      {\cal L} & = & \bar\psi [ \gamma ( i\partial - g_\omega\omega ) - m_{\rm nuc} - g_\sigma \sigma ] \psi 
                                                                                            \nonumber \\
               & + & {1 \over 2} (\partial \sigma)^2 - U(\sigma)
                 -  {1 \over 4} F_{\mu\nu} F^{\mu\nu} + {1 \over 2} m_\omega^2 \omega^2~,    \label{Lagr}
\end{eqnarray}
where $\psi$ is the nucleon field, $\sigma$ and $\omega$ are the scalar-isoscalar and the vector-isoscalar
meson fields, respectively; $F_{\mu\nu} = \partial_\mu \omega_\nu - \partial_\nu \omega_\mu$.
We neglect the isovector meson and the electromagnetic contributions. The Lagrangian density (\ref{Lagr})
contains the nonlinear self-interactions of the $\sigma$-field:
\begin{equation}
      U(\sigma) = {1 \over 2} m_\sigma^2 \sigma^2 +  {1 \over 3} g_2 \sigma^3 
                                                  +  {1 \over 4} g_3 \sigma^4~.              \label{U}
\end{equation}
The Lagrange's equations of motion for the nucleon, $\sigma$- and $\omega$-fields are written as follows:
\begin{eqnarray}
     & & [ \gamma ( i\partial - g_\omega\omega ) - m_{\rm nuc} - g_\sigma \sigma ] \psi =  0~,     \label{DiracEq} \\
     & & \partial_\mu\partial^\mu \sigma + { \partial U(\sigma) \over \partial \sigma } 
                                                                            =  -g_\sigma \bar\psi \psi~,
                                                                                          \label{KGeqSigma} \\
     & & \partial_\mu F^{\mu\nu} + m_\omega^2 \omega^\nu = g_\omega \bar\psi \gamma^\nu \psi~. 
                                                                                           \label{KGeqOmega}
\end{eqnarray}
The $\sigma$- and $\omega$-fields are treated as classical ones, i.e. we replace these fields
by their expectation values in Eqs.(\ref{DiracEq}, \ref{KGeqSigma}, \ref{KGeqOmega}). Assuming that
the meson fields are varying much more slowly in time and space with respect to the nucleon field, we 
consider the plane-wave solutions of Eq.(\ref{DiracEq}):
\begin{equation}
                  \psi^{(\pm)} \propto \exp(\mp ipx)~,                          \label{planeWave}
\end{equation}
where $x \equiv (t,{\bf r})$, $p \equiv (p^0,{\bf p})$, and the upper (lower) sign corresponds to the nucleon 
(antinucleon). The dispersion relation is then obtained from (\ref{DiracEq}):
\begin{equation}
                 p^0 = \pm g_\omega \omega^0 
                     + \sqrt{ ({\bf p} 
                               \mp  g_\omega\mbox{\boldmath ${\mathbf \omega}$ \unboldmath})^2 
                                + (m_{\rm nuc}^\star)^2 }~,                                       \label{dispRel}
\end{equation}
where
\begin{equation}
                m_{\rm nuc}^\star = m_{\rm nuc} + g_\sigma \sigma                                         \label{meff}
\end{equation}
is the nucleon effective (Dirac) mass.

The distribution function $f(x,{\bf p})$ of a given particle species in the phase space
$({\bf r},{\bf p})$ is defined now such that 
$f(x,{\bf p}) \frac{g d^3r d^3p}{(2\pi)^3}$ = (number of particles of that species in the phase 
space element $d^3r d^3p$), where $g=4$ is the spin-isospin degeneracy. 
The space-time evolution of the (anti)nucleon phase space distribution function is described by the BUU 
equation
\begin{equation}
   { \partial f \over \partial t} 
 + { \partial p^0 \over \partial {\bf p} } { \partial f \over \partial {\bf r} }
 - { \partial p^0 \over \partial {\bf r} } { \partial f \over \partial {\bf p} }
 = I_{\rm coll}[f]~,                                                       \label{BUU}
\end{equation}
where --- in spirit of the Fermi liquid theory \cite{Landau,Pomeranchuk} --- 
the single-particle energy (\ref{dispRel}) is used as a one-body Hamiltonian function.
The l.h.s. of Eq.(\ref{BUU}) describes the propagation of particles in the mean field, while the
r.h.s. is the collision integral. The explicit form of $I_{\rm coll}$ for elastic scattering 
is given below (see Eqs.(\ref{Icoll_exp}),(\ref{Icoll}),(\ref{I3el_exp})).

It is convenient to perform a variable transformation in Eq.(\ref{BUU}) by using the kinetic four-momentum
\begin{equation}
   p^{\star\mu} = p^\mu \mp g_\omega \omega^\mu~.                           \label{pKin}
\end{equation}
The distribution function $f^\star(x,{\bf p^\star})$ in the kinetic phase space $({\bf r},{\bf p^\star})$
is defined such that $f^\star(x,{\bf p^\star}) \frac{g d^3r d^3p^\star}{(2\pi)^3}$ = (number of particles 
in the kinetic phase space element $d^3r d^3p^\star$). Since $d^3p=d^3p^\star$, which is valid for 
the momentum-independent $\omega$-field, one gets:
\begin{equation}
   f^\star(x,{\bf p^\star})=f(x,{\bf p})~.                                     \label{fStar}
\end{equation}
Expressing the l.h.s. of Eq.(\ref{BUU}) in terms of the kinetic quantities we obtain the transport equation
(c.f. Ref. \cite{Blaettel93}):
\begin{equation}
  (p_0^\star)^{-1}
  [ p_\mu^\star \partial_x^\mu + (\pm g_\omega p_\nu^\star F^{\alpha\nu} 
                                   + m_{\rm nuc}^\star(\partial_x^\alpha  m_{\rm nuc}^\star))
    \partial^{p^\star}_\alpha ] f^\star(x,{\bf p^\star}) = I_{\rm coll}[f^\star]~,       \label{BUUstar}
\end{equation}
where $\alpha=1,2,3$, and $p_0^\star$ is determined from the mass shell condition
$(p^\star)^2 = (m_{\rm nuc}^\star)^2$.

The collision integral, generally, can be expanded in the number of colliding particles
(c.f. Ref. \cite{BRV92} and refs. therein):
\begin{equation}
   I_{\rm coll}[f^\star] =  I_{\rm coll, 2b}[f^\star] + I_{\rm coll, 3b}[f^\star] %
                          + I_{\rm coll, 4b}[f^\star] + ...~.                              \label{Icoll_exp}
\end{equation}
We will restrict ourselves to the first two terms in (\ref{Icoll_exp}) only.
It is straightforward  to write down the elastic contributions to $I_{\rm coll, 2b}$ and 
$I_{\rm coll, 3b}$ in the case of identical fermions: 
\begin{eqnarray}
&& I_{\rm coll, 2b}^{\rm elastic}[f_1^\star] = \frac{1}{2!} \, \frac{m_1^\star}{p_1^{\star0}}
   \int\, \frac{ g d^3 p_{1^\prime}^\star m_{1^\prime}^\star }{ (2\pi)^3 p_{1^\prime}^{\star0} } %
   \int\, \frac{ g d^3 p_{2^\prime}^\star m_{2^\prime}^\star }{ (2\pi)^3 p_{2^\prime}^{\star0} } % 
   \int\, \frac{ g d^3 p_2^\star m_2^\star }{ (2\pi)^3 p_2^{\star0} }                           \nonumber \\
&&  \times \overline{|M|^2_{12 \to 1^\prime2^\prime}} \,
   (2\pi)^4 \, \delta^{(4)}(p_1+p_2-p_{1^\prime}-p_{2^\prime}) \, 
   (  f_{1^\prime}^\star f_{2^\prime}^\star \bar{f}_1^\star \bar{f}_2^\star %
    - f_1^\star f_2^\star \bar{f}_{1^\prime}^\star \bar{f}_{2^\prime}^\star )~,                \label{I2el}
\end{eqnarray}
\begin{eqnarray}
&& I_{\rm coll, 3b}^{\rm elastic}[f_1^\star] = \frac{1}{3!\,2!} \, \frac{m_1^\star}{p_1^{\star0}}
   \int\, \frac{ g d^3 p_{1^\prime}^\star m_{1^\prime}^\star }{ (2\pi)^3 p_{1^\prime}^{\star0} } %
   \int\, \frac{ g d^3 p_{2^\prime}^\star m_{2^\prime}^\star }{ (2\pi)^3 p_{2^\prime}^{\star0} } %
   \int\, \frac{ g d^3 p_{3^\prime}^\star m_{3^\prime}^\star }{ (2\pi)^3 p_{3^\prime}^{\star0} }    \nonumber \\
&& \times \int\, \frac{ g d^3 p_2^\star m_2^\star }{ (2\pi)^3 p_2^{\star0} }   %
            \int\, \frac{ g d^3 p_3^\star m_3^\star }{ (2\pi)^3 p_3^{\star0} }\,   %
            \overline{|M|^2_{123 \to 1^\prime2^\prime3^\prime}}      %
            (2\pi)^4 \, \delta^{(4)}(p_1+p_2+p_3-p_{1^\prime}-p_{2^\prime}-p_{3^\prime})    \nonumber \\      
&& \times (  f_{1^\prime}^\star f_{2^\prime}^\star  f_{3^\prime}^\star \bar{f}_1^\star \bar{f}_2^\star \bar{f}_3^\star  %
             - f_1^\star f_2^\star f_3^\star \bar{f}_{1^\prime}^\star \bar{f}_{2^\prime}^\star \bar{f}_{3^\prime}^\star )~,
                                                                                            \label{I3el}
\end{eqnarray}
where $\overline{|M|^2_{12 \to 1^\prime2^\prime}}$ and 
$\overline{|M|^2_{123 \to 1^\prime2^\prime3^\prime}}$ are the spin- and isospin-averaged over initial
and final states matrix elements squared for two- and three-body collisions, respectively;
$f_i^\star \equiv f^\star(x,{\bf p_i^\star})$, $\bar{f}_i^\star \equiv 1 - f_i^\star$  
($i=1,2,3,1^\prime,2^\prime,3^\prime$). 
The matrix elements satisfy the detailed balance relations
\begin{eqnarray} 
\overline{|M|^2_{1^\prime2^\prime \to 12}} &=& \overline{|M|^2_{12 \to 1^\prime2^\prime}}~,         \\ 
\overline{|M|^2_{1^\prime2^\prime3^\prime \to 123}} &=& \overline{|M|^2_{123 \to 1^\prime2^\prime3^\prime}}~,
                                                                                            \label{detbal}
\end{eqnarray}
which are used in Eqs.(\ref{I2el}),(\ref{I3el}). The normalization of Bjorken and Drell \cite{BD} is
used for the matrix elements, which leads to the appearance of the fermion Dirac masses in 
Eqs. (\ref{I2el}),(\ref{I3el}). In the case of the momentum-independent $\sigma$-field used in this work, 
we have for the identical fermions 
$m_1^\star=m_2^\star=m_3^\star=m_{1^\prime}^\star=m_{2^\prime}^\star=m_{3^\prime}^\star$.
However, in Eqs.(\ref{I2el}),(\ref{I3el}) and everywhere below in the expressions for the collision integrals, 
we keep, for clarity, different subscripts for the Dirac masses of different particles.

Introducing the in-medium elastic differential scattering cross section
\begin{eqnarray}
d\sigma_{1 2 \to 1^\prime 2^\prime}^\star %
&=&       (2\pi)^4 \, \delta^{(4)}(p_1+p_2-p_{1^\prime}-p_{2^\prime})\, %
         \overline{|M|^2_{12 \to 1^\prime2^\prime}}    \nonumber \\
&\times& \frac{m_1^\star m_2^\star}{I_{12}^\star}\, %  
         \frac{ g d^3 p_{1^\prime}^\star m_{1^\prime}^\star }{ (2\pi)^3 p_{1^\prime}^{\star0} }\, %
         \frac{ g d^3 p_{2^\prime}^\star m_{2^\prime}^\star }{ (2\pi)^3 p_{2^\prime}^{\star0} }\, % 
         \frac{1}{2!}                                                                        \label{dSigmaStar}
\end{eqnarray}
with 
\begin{equation}
   I_{12}^\star \equiv \sqrt{(p_1^\star p_2^\star)^2 - (m_1^\star m_2^\star)^2}          \label{I12Star}
\end{equation}
being the in-medium flux factor, one can rewrite the two-body elastic collision term as follows:
\begin{equation}
I_{\rm coll, 2b}^{\rm elastic}[f_1^\star] =\int\, \frac{ g d^3 p_2^\star }{ (2\pi)^3 } %
                                           \int\, d\sigma_{1 2 \to 1^\prime 2^\prime}^\star %
                                           \, v_{12}^\star % 
   (  f_{1^\prime}^\star f_{2^\prime}^\star \bar{f}_1^\star \bar{f}_2^\star %
    - f_1^\star f_2^\star \bar{f}_{1^\prime}^\star \bar{f}_{2^\prime}^\star )~,                \label{Icoll}
\end{equation}
where 
\begin{equation}
    v_{12}^\star = I_{12}^\star / (p_1^{\star 0} p_2^{\star 0})                              \label{v12Star}
\end{equation}
is the relative velocity of colliding particles.

In agreement with the low-energy in-medium calculations of the elastic NN
scattering cross section (c.f. Ref. \cite{PP92}), one can  
neglect the medium dependence of the matrix element $\overline{|M|^2_{12 \to 1^\prime2^\prime}}$ approximately.
This approximation will be used in the next
subsection in order to evaluate the in-medium baryon-baryon cross sections (c.f. Eq.(\ref{sigMed}) below).
 
To solve Eq.(\ref{BUUstar}), the distribution function in the kinetic phase space is projected onto
test particles:
\begin{equation}
   f^\star(x,{\bf p^\star}) = \frac{ (2\pi)^3 }{ g N } \sum_{i=1}^{A N} \delta( {\bf r} - {\bf r_i}(t) )
                                                         \delta( {\bf p^\star} - {\bf p_i^\star}(t) )~, 
                                                                                \label{testPart}
\end{equation}
where $A$ is the number of nucleons and $N$ is the number of test particles per nucleon.
The centroids of the $\delta$-functions are evolving in time between the two- or three-body collisions
according to the following equations, which can be obtained by substituting (\ref{testPart}) into
(\ref{BUUstar}) and putting the collision term equal to zero (c.f. Refs. \cite{LCMW92,Blaettel93}):
\begin{eqnarray}
& & {d {\bf r}_i \over d t} =  { {\bf p}^\star_i \over p_i^{\star 0} }~,                       \label{rDot} \\
& & {d p_i^{\star \alpha} \over d t } = \pm g_\omega { p_{i\nu}^\star \over p_i^{\star 0} } F^{\alpha\nu}
                              + { m_{\rm nuc}^\star \over p_i^{\star 0} } \partial_x^\alpha m_{\rm nuc}^\star     
                                                                                              \label{pStarDot}
\end{eqnarray}
with $\alpha=1,2,3$ and $\nu=0,1,2,3$.

The $\sigma$- and $\omega$-fields are calculated from the Klein-Gordon equations 
(\ref{KGeqSigma}),(\ref{KGeqOmega}) by neglecting the derivatives of the fields in space and time:
\begin{eqnarray}
& & m_\sigma^2 \sigma + g_2 \sigma^2 + g_3 \sigma^3 = - g_\sigma \rho_s~,           \label{eqSigma}\\
& & m_\omega^2 \omega^\nu = g_\omega j_B^\nu~.                                      \label{eqOmega}\\
\end{eqnarray}
The scalar density $\rho_s(x)=<\bar \psi(x) \psi(x)>$ and the baryon current
$j_B^\nu(x)=<\bar \psi(x) \gamma^\nu \psi(x)>$ are expressed via the
(anti)nucleon phase space distribution functions:
\begin{eqnarray}
\rho_s(x) & = & \int\, \frac{ g d^3 p^\star m_{\rm nuc}^\star }{ (2\pi)^3 p^{\star0} } %
               (  f^\star_{\rm nucleon}(x,{\bf p^\star}) %
                + f^\star_{\rm antinucleon}(x,{\bf p^\star}) )~,                       \label{rhoScalar} \\
j_B^\nu(x) & = & \int\, \frac{ g d^3 p^\star p^{\star\nu} }{ (2\pi)^3 p^{\star0} } %
               (  f^\star_{\rm nucleon}(x,{\bf p^\star}) %   
                - f^\star_{\rm antinucleon}(x,{\bf p^\star}) )~.                       \label{jBaryon}
\end{eqnarray}
Since we use the distribution functions in the kinetic phase space, Eqs.(\ref{eqSigma}),(\ref{rhoScalar}) do
not depend explicitly on the vector field $\omega$. This simplifies the selfconsistent numerical calculation 
of the meson mean fields strongly.

We assume, for simplicity, the same coupling constants of all other baryons with the $\sigma$- and 
$\omega$-fields as for the nucleon. Correspondingly, in actual calculations of the scalar density
and of the baryon current, the partial contributions from all the baryons present in the system 
are taken into account. The baryon test particles are propagated according to Eqs.(\ref{rDot}),
(\ref{pStarDot}) with a replacement $m_{\rm nuc}^\star \to m_B^\star = m_B + g_\sigma \sigma$, where $m_B$ is 
the vacuum mass of the baryon $B$. The potentials acting on mesons are neglected. Thus, the mesons are 
propagated freely between the two- or three-body collisions.

The numerical values of the RMF parameters were chosen according to the parameter set NL2 from Ref. 
\cite{LCMW92}: $m_\sigma=550.5$ MeV, $m_\omega=783.3$ MeV, $g_\sigma=8.50$, $g_\omega=7.54$, 
$g_2=-50.37$ fm$^{-1}$, $g_3=-6.26$. This parameter set \cite{com1} produces the incompressibility 
$K=210$ MeV and the ratio $m_{\rm nuc}^\star/m_{\rm nuc}=0.83$ at normal nuclear matter density. At high
densities, the NL2 parameter set gives a rather soft equation of state \cite{Blaettel93}.
This agrees with the BEVALAC data on collective flow in HIC at about 1 A GeV beam energy
\cite{Blaettel93}.

The $\sigma$-field and the baryon four-current have been computed on the space grid 
with the cell sizes $\simeq (1 \times 1 \times 1/\gamma)$ fm covering the collision zone.
The smaller cell size in the longitudinal direction is needed in order to resolve 
the density profiles of the colliding nuclei, which are Lorentz contracted by the 
$\gamma$-factor in the center-of-mass (c.m.) frame. We have used the parallel ensemble technique 
(c.f. Ref. \cite{BDG88}) with $N=200$ test particles per nucleon. 
To get smooth meson mean fields, the $\delta$-function in coordinate space 
$\delta( {\bf r} - {\bf r_i}(t) )$ in Eq. (\ref{testPart}) 
has been replaced by the Lorentz contracted gaussian
\begin{equation}
        \rho_i({\bf r}) = {\gamma \over (2\pi)^{3/2} L^3} 
          \exp\left\{ -{(x-x_i(t))^2 \over 2 L^2} -{(y-y_i(t))^2 \over 2 L^2}
                      -{(z-z_i(t))^2 \gamma^2 \over 2 L^2} \right\}            \label{gaussian}
\end{equation}
with $L \simeq 1$ fm. The equations of motion (\ref{rDot}),(\ref{pStarDot}) for the test 
particles have been solved by using the $O(\Delta t^2)$ predictor-corrector method.
The space derivatives in the r.h.s. of Eq.(\ref{pStarDot}) have been computed using the
central differences, which produces a second order accuracy also in space.
The numerical scheme conserves energy with an accuracy better than 3\% of the c.m. kinetic energy
for the studied reactions.

\subsection{In-medium cross sections}

The most important part of the GiBUU model is the collision integral which includes the hadron-hadron cross 
sections. In the case of cascade calculation, neglecting any mean field effects, the vacuum cross sections 
are invoked.
These cross sections are based either on the resonance model or on the phenomenological parameterizations of
the experimental data with some reasonable extrapolations to the not-measurable channels (e.g. in the 
case of meson-meson collisions).   
The detailed description of the low energy resonance cross sections is given in Refs. 
\cite{Teis97,EBM99,MEPhD}.
The high energy cross section parameterizations --- used for the FRITIOF event generator --- are explained 
in Ref. \cite{Falter04}. For the meson-meson cross sections, we refer the reader to Ref. \cite{WLM05}.

Using the RMF model in particle propagation requires also to introduce in-medium modifications 
of the cross sections.
This should be already clear, since the particle production thresholds include now the Dirac masses instead of
the vacuum masses.
To evaluate the cross sections in the case of calculations with RMF, we will apply two 
different schemes.

In the first scheme, a so-called corrected invariant energy of the two colliding particles 1 and 2
is computed as
\begin{equation}
        \sqrt{s_{\rm corr}} = \sqrt{s^\star} - (m_1^\star - m_1) - (m_2^\star - m_2)~,            
                                                                                 \label{srtsCorr}
\end{equation}
where 
\begin{equation}
  s^\star = (p_1^\star + p_2^\star)^2~.                                          \label{sStar}
\end{equation}
The corrected invariant energy is an analog of the vacuum 
invariant energy, since the scalar selfenergies of the colliding particles are subtracted in the r.h.s. of
Eq.(\ref{srtsCorr}). The quantity $\sqrt{s_{\rm corr}}$ is then used in calculation of any reaction
cross section $\sigma_{12 \to X}(\sqrt{s_{\rm corr}})$. Due to the same scalar selfenergies
for all the baryons, this scheme ensures the correct threshold conditions for all binary processes
except for $B \bar B$ production/annihilation. The last processes are, however, not important
at the beam energies considered in the present work.

In the second scheme, we follow the approach of Ref. \cite{WLM05}.
There the in-medium cross section of a process
$B_1 B_2 \to B_3 B_4 M_5 M_6 ... M_N$ 
with $B_{1,2}$ and $B_{3,4}$ as incoming and outgoing baryons and
$M_{5,\dots,N}$ as produced mesons, is expressed in a form:
\begin{equation}
    \sigma^{med}(\sqrt{s^\star}) = F \sigma^{vac}(\sqrt{s_{\rm corr}})~.       \label{sigMed}
\end{equation}
The modification factor $F$ is
\begin{equation}
    F \equiv {m_1^\star m_2^\star m_3^\star m_4^\star \over m_1 m_2 m_3 m_4}
             {I \over I^\star}
{\Phi_{N-2}(\sqrt{s^\star};m_3^\star,m_4^\star,...,m_N^\star)  \over
 \Phi_{N-2}(\sqrt{s_{\rm corr}};m_3,m_4,...,m_N)}~,                         \label{F}
\end{equation}
where
\begin{equation}
   \Phi_n( M; m_1, m_2, ..., m_n ) = \int\, d\Phi_n(P; p_1, p_2, ..., p_n)    \label{Phi}
\end{equation}
is the $n$-body phase space volume with $m_i^2 = p_i^2$ ($i=1,2,...,n$) and $M^2 = P^2$, 
\begin{equation}
   d\Phi_n(P; p_1, p_2, ..., p_n) = \delta^{(4)}( P - p_1 - p_2 - ... - p_n ) %
               \frac{ d^3p_1 }{ (2\pi)^3 2p_1^0 } %
               \frac{ d^3p_2 }{ (2\pi)^3 2p_2^0 } %                  
             \cdots \frac{ d^3p_n }{ (2\pi)^3 2p_n^0 }                      \label{dPhi}
\end{equation}
is the element of the $n$-body phase space volume (c.f. Ref. \cite{PDG02});
\begin{eqnarray}
   I_{12}   & = & q( \sqrt{s_{\rm corr}}, m_1, m_2 ) \sqrt{s_{\rm corr}}~,                 \label{I} \\
   I_{12}^\star & = & q( \sqrt{s^\star}, m_1^\star, m_2^\star ) \sqrt{s^\star}             \label{Istar}
\end{eqnarray}
are the vacuum and in-medium flux factors (c.f. Eq.(\ref{I12Star})) with
\begin{equation}
    q( \sqrt{s}, m_1, m_2 )
= \sqrt{ ( s + m_1^2 - m_2^2 )^2 / (4s) - m_1^2 }                                      \label{q}
\end{equation}
being the center-of-mass (c.m.) momentum. Eqs.(\ref{sigMed}),(\ref{F}) take into account the 
in-medium modification of the Dirac plane wave normalization (a factor of 
$m_1^\star m_2^\star m_3^\star m_4^\star / (m_1 m_2 m_3 m_4)$), of the flux factor and of the phase space 
volume. However, it is assumed that the matrix element of the reaction is not in-medium modified. Due to 
the Dirac mass reduction with the baryon density, the in-medium baryon-baryon cross section is substantially 
reduced in nuclear medium. In the present work, we have extended the method of Ref. \cite{WLM05} by applying
Eqs.(\ref{sigMed}),(\ref{F}) also for the in-medium modification of the channel $B_1 B_2 \to B_3 B_4$,
i.e. for the elastic baryon-baryon collisions or resonance excitation in baryon-baryon collisions without
outgoing mesons. 

Meson-baryon cross sections were kept always as the vacuum ones. We believe that this is
a reasonable assumption, since the modification factor will be proportional to $(m^\star)^2$ in this case, 
while the modification factor for the baryon-baryon cross sections is proportional to $(m^\star)^4$ 
(c.f. Eq.(\ref{F})).
Thus, the meson-baryon cross sections are less subject to the in-medium modifications. On the other
hand, the implementation of the in-medium meson-baryon cross sections is more difficult, since
in this case also the resonance decay widths should be consistently modified to preserve detailed
balance in the channel $MB \leftrightarrow R$.

For brevity, the first scheme will be refered to as the calculation with vacuum cross sections below.
The second scheme will be called the calculation with the in-medium cross sections.

\subsection{Three-body collisions}

Little is known about the three-body forces even in ground state nuclear matter.
The problem of particle production in three-body collisions is even harder. There
are no experimental data on this subject and the corresponding matrix elements
are not obtainable from data \cite{com2}.
Thus, in simulating the three-body collisions, we apply a simple geometrical method 
similar to that of Refs. \cite{Mrow85,BRV92,BM93,BGM94}. For the details of derivation we
refer the reader to Ref. \cite{BGM94}.   

The geometrical method of Refs. \cite{Mrow85,BRV92,BM93,BGM94} is based on the hard-sphere 
collision picture, i.e. the potential acting between colliding nucleons 1 and 2 is assumed 
to be infinitely repulsive at relative distances $d_{12} \leq R_{\rm max}$
and zero at $d_{12} > R_{\rm max}$, where $R_{\rm max}$ is equal to the sum of the matter radii 
of the colliding particles. The quantity $R_{\rm max}$ can be expressed in terms of 
the total (in-medium) interaction cross section $\sigma_{12}^\star$ of the particles 1 and 2 
as follows:
\begin{equation}
   R_{\rm max} = \sqrt{\sigma_{12}^\star/\pi}~.                         \label{Rmax}
\end{equation}

The three-body collision is assumed to happen, if (i) the two nucleons 1 and 2 are about to 
collide according to the geometrical collision criterion \cite{BDG88}
and (ii) the third particle is found in a sphere of radius $R_{\rm max}$ centered at the collision
point of 1 and 2. The geometrical collision criterion for selecting the colliding pair 1 and 2 means
that these particles approach their minimum separation distance during the given time step, 
and this distance is less than $R_{\rm max}$. Thus, the three-body collision takes place
when all the three particles are found simultaneously in the interaction volume which is 
here the sphere of radius $R_{\rm max}$ centered at the c.m. of 1 and 2 in coordinate space.

The hard-sphere collision picture reflects, in a natural way, the short-range character of the nuclear 
forces acting between hadrons. However, in a high-energy nucleus-nucleus collision case, one
must modify this picture to account for the Lorentz contraction of the matter radii of the colliding 
hadrons. The relativistic gas of the Lorentz-contracted hard spheres in thermal equilibrium has been 
already considered in Refs. \cite{BGSG00,Bugaev07} in order to derive the Lorentz corrections to the 
excluded volume in the Van-der-Waalls equation of state.

Therefore, in the present work we define the interaction volume as an axially symmetric ellipsoid 
contracted along the collision axis of 1 and 2 by the average $\gamma$-factor in their c.m. frame  
\begin{equation}
   \gamma_{12}=((\gamma_1+\gamma_2)/2)_{\rm cm12}~,                                   \label{gamma12}
\end{equation}
where $\gamma_1=p_1^{\star0}/m_1^\star$ and $\gamma_2=p_2^{\star0}/m_2^\star$. The ellipsoid has a half-axis 
of $R_{\rm max}/\gamma_{12}$ along the particle 1 momentum in the c.m. frame of 1 and 2 and a half-axis of 
$R_{\rm max}$ in a transverse direction.

We now select the set of all particles inside the ellipsoid which are different from
1 and 2. In principle, all the particles from the set plus the colliding pair form 
a many-body colliding system. However, for technical reasons, we restricted ourselves in
this work to a simulation of three-body collisions only. Therefore, we choose only one 
particle from the set, namely, the closest particle to the c.m. of 1 and 2 --- as a 
participant of a three-body collision. We will denote this particle as 3 below.

Next we simulate the actual three-body collision event of the triple
1, 2 and 3. We denote the initial kinetic momenta of the triple as ${\bf p_1^\star}$, ${\bf p_2^\star}$
and ${\bf p_3^\star}$.  
Following Ref. \cite{BRV92}, the momenta of the triple are, first, redistributed
microcanonically. This is done by sampling the new kinetic momenta ${\bf p_{1^\prime}^\star}$,
${\bf p_{2^\prime}^\star}$ and ${\bf p_{3^\prime}^\star}$  according to the probability
\begin{equation}
   d{\cal P} \propto d\Phi_3(p_1^\star+p_2^\star+p_3^\star;%
                           \, p_{1^\prime}^\star,\, p_{2^\prime}^\star,\, %
                           p_{3^\prime}^\star)~,                         \label{Prob}
\end{equation}
where $d\Phi_3$ is the three-body phase space volume element (Eq.(\ref{dPhi})).
It is assumed here that the particles 1,2 and 3 keep their identity. In particular, they stay
on their initial Dirac mass shells: $(p_{i^\prime}^\star)^2=(p_i^\star)^2=(m_i^\star)^2$, i=1,2,3.

After the redistribution of the momenta, the two-body collision of the particles 1 and 2 with
four-momenta $p_{1^\prime}^\star$ and $p_{2^\prime}^\star$ is simulated in a usual way. This can lead
to either elastic or inelastic scattering, including multiple particle production through
the FRITIOF mechanism. In Appendix A, the formal expressions for the three-body collision integral
and the corresponding matrix element squared are given which reflects the procedure discussed above
in the case of elastic collisions of identical fermions. 

In the actual numerical simulations, we performed the search for the third particle, which can
be either a baryon or a meson, only if the particles 1 and 2 are both baryons or a meson and a baryon. 
For the meson-meson collisions, the search for the third particle has not been done. The meson-meson 
collisions are, however, relatively soft and can not influence much the high-$p_t$ part of the 
spectra, which are of the primary interest in our present study.     

To save the CPU time, we have also switched-off the Pauli blocking of the final state in all collision 
and resonance decay processes. We have checked by direct calculation, that above $E_{\rm lab}=2$ A GeV the 
Pauli blocking is negligibly small.

\section{Numerical results}

We have performed the calculations for central Au+Au collisions at beam energies of 2-20 A GeV
and central Pb+Pb collisions at 30 and 40 A GeV. The time evolution of the systems in the c.m. frame of
the colliding nuclei has been followed up to 30 fm/c using a variable time step. The size of the time
step was adjusted to reduce the spurious effect of multiple scatterings of the same particle within 
the given time step. 
 
In order to see an influence of the various physical ingredients of our model,
four types of calculation have been done: (i) pure binary cascade calculation without mean field and 
using the vacuum cross sections;
(ii) the calculation with the RMF, with only binary collisions and vacuum cross sections;
(iii) the same as (ii) plus the three-body collisions;
(iv) the same as (iii), but with the in-medium baryon-baryon cross sections.

It is necessary to point out, that when calculating the radius $R_{\rm max}$ of Eq.(\ref{Rmax})
we have used the cross section $\sigma_{12}^\star$ somewhat different from the actual total two-body 
cross section implemented in the model. Namely, in calculations with vacuum cross 
sections (i),(ii) and (iii), $\sigma_{12}^\star$ was set to 40 mb for a baryon-baryon collision, 
which is an asymptotic high energy value of the total $pp$ cross section. In the case of the calculation 
with in-medium baryon-baryon cross sections (iv), we used for $\sigma_{12}^\star$ the in-medium $pp$ 
cross section determined by summation of all possible final state channels. The in-medium cross section of 
each final channel was obtained according to Eq.(\ref{sigMed}). For the meson-baryon collisions, the constant 
cross section $\sigma_{12}^\star=20$ mb was always used in Eq.(\ref{Rmax}). This value is close to the asymptotic
high energy $\pi^+ p$ total cross section. The rescaling factors due to the leading quark numbers
(c.f. Sect. II) have not been taken into account in the calculation of the interaction volume. 
To avoid misunderstanding, we stress that these simplifying assumptions have been made when
calculating the interaction volume only, but not in determination of the two-body collision
partners by the geometrical collision criterion. 

\subsection{ Time evolution }

\begin{figure}

\includegraphics{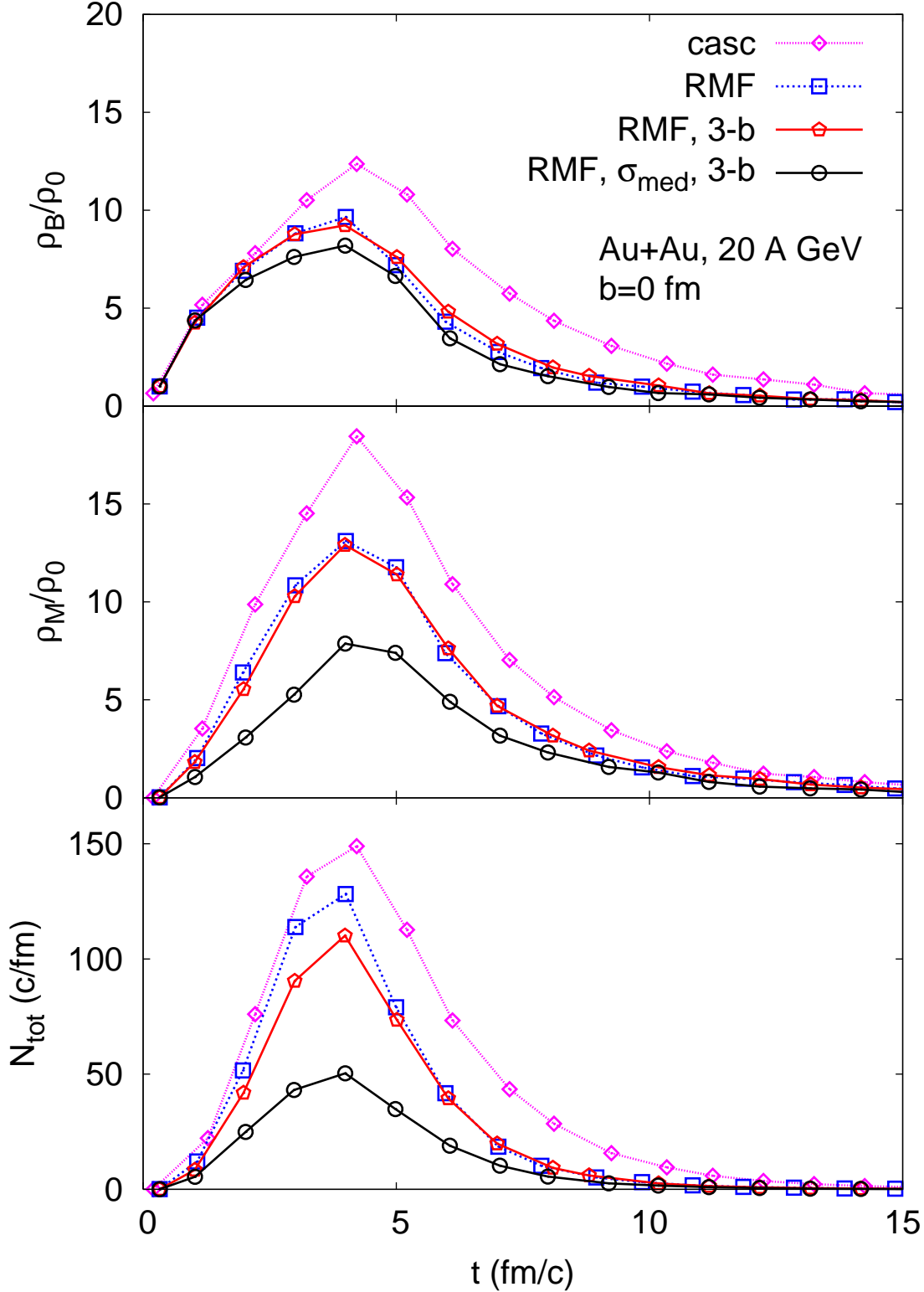}

\vspace*{1cm}

\caption{\label{fig:freq_tot} (color online) Time evolution of the central baryon density (top panel),
of the central meson density (middle panel) and of the total collision frequency (bottom panel) for  
central Au+Au collision at 20 A GeV. Binary cascade calculation is represented by dotted lines with open rombuses.
RMF calculation including binary collisions only and vacuum cross sections is shown by dashed lines with open boxes.
Results with RMF including also the three-body collisions with vacuum cross sections are plotted by solid lines with open 
pentagons. RMF calculation with the three-body collisions and in-medium cross sections is shown by solid lines with open 
circles.}

\end{figure}

Fig.~\ref{fig:freq_tot} shows the time dependencies of the central baryon and meson densities and
of the total collision frequency for the Au+Au system at 20 A GeV and b=0 fm. The baryon and meson densities
have been computed in the central 1fm$\times$1fm$\times$1fm cube. The total collision frequency $N_{\rm tot}$ 
includes the two- and (when switched-on) the three-body collisions and has been determined in the larger 
3fm$\times$3fm$\times$3fm central cube to reduce the statistical fluctuations. The quantity $N_{\rm tot}$ 
reaches its maximum at about 4 fm/c, when also the central baryon and meson densities are maximal, and 
drops rapidly later on. We observe, that the RMF reduces the maximum baryon and meson densities 
and also leads to a faster expansion of the compressed system. The three-body collisions do not influence 
the central densities and influence the total collision frequency only weakly. The weakness of the dependence 
of the total collision frequency on the three-body collisions is due to the fact, that the inclusion of the 
three-body collisions reduces the number of the pure two-body collisions, i.e. collisions where there is no 
other particles in the interaction volume of the primary colliding pair 1 and 2.
Indeed only about 10-20\% of all collision events are now pure two-body (c.f. Fig.~\ref{fig:freq} below).

It is interesting, that the inclusion of three-body collisions even reduces $N_{\rm tot}$ slightly. 
This happens, since the particle $3^\prime$ and the particles emerging from the interaction of $1^\prime$ 
and $2^\prime$ are final state particles of a three-body collision event (see Sect. IIC). The final state 
particles are not allowed to rescatter on each other before at least one of them will rescatter on another 
particle. This leads to an overall reduction of the total collision frequency, when the three-body collisions 
are included. Indeed, without the three-body collisions, the particle 3 would be allowed to rescatter on the 
collision products of 1 and 2.
 
Finally, using the in-medium baryon-baryon cross sections influences the baryon density rather weakly. 
However, the meson production and the total collision frequency are strongly decreased in this case.

Fig.~\ref{fig:freq} presents the ratio of the frequency of three-body collisions to the total collision 
frequency (two- plus three-body) as a function of time for the Au+Au central collision at 20 A GeV 
(bottom left and right panels). In order to understand this ratio better, we also show in Fig.~\ref{fig:freq} 
the time dependencies of the central baryon and meson densities along with their sum and the average 
$\gamma$-factor of the two-body collisions (top left and right panels). 
The cross section $\sigma_{12}^\star$ used to compute the radius $R_{\rm max}$ of Eq.(\ref{Rmax}) averaged
over colliding pairs is also plotted vs time in Fig.~\ref{fig:freq} (middle left and right panels).
Calculations both with vacuum (left column) and in-medium (right column) cross sections are shown in 
Fig.~\ref{fig:freq}. We see that the ratio $N_3/N_{\rm tot}$ reaches the maximum value of 0.9 for the vacuum 
cross sections and 0.8 for the in-medium cross sections. 

The ratio $N_3/N_{\rm tot}$ can be estimated on the basis of the Poissonian distribution for the probability 
to find $n=0,1,2,...$ particles in the interaction volume of the colliding pair 1 and 2 
(c.f. Ref. \cite{LGM06}):
\begin{equation}
   P_n = \frac{\lambda^n}{n!} \exp(-\lambda)~,                        \label{Pn}
\end{equation}
where $\lambda=\rho_{\rm tot} <V_{12}>$ with $<V_{12}>$ being the averaged interaction volume 
(c.f. Eq.(\ref{V12})):
\begin{equation}
   <V_{12}> \equiv \frac{4}{3} \pi \left(\frac{<\sigma_{12}^\star>}{\pi}\right)^{3/2}  %
            <\gamma_{12}>^{-1}~.                                           \label{V12_aver}
\end{equation}
For the meson-free matter, the quantity $\lambda$ is proportional to the gas parameter (Eq.(\ref{gammaGas})).
Since, by definition (see Sect. IIC), the three-body collision happens if $n \geq 1$, the following 
estimate can be done:
\begin{equation}
   N_3/N_{\rm tot} \simeq 1 - P_0 = 1 - \exp(-\lambda)~.        \label{estimate}
\end{equation}
In the nonrelativistic limit for the meson-free matter with $\lambda \ll 1$, Eq.(\ref{estimate}) is identical 
to the result of Ref. \cite{Mrow85}. The estimate (\ref{estimate}) is depicted by the dashed lines in the 
bottom left and right panels of Fig.~\ref{fig:freq}. We observe, that Eq.(\ref{estimate}) reproduces the 
overall behaviour of the directly computed ratio $N_3/N_{\rm tot}$. In particular, one can see, that the flat 
maximum of $N_3/N_{\rm tot}$ calculated with the vacuum cross sections is caused by larger 
$<\sigma_{12}^\star>$ at the initial stage of collision. In the case of the in-medium baryon-baryon cross 
sections, the value of $<\sigma_{12}^\star>$ drops quickly at the beginning, reaching the minimum at about 
2 fm/c. This reduces the ratio $N_3/N_{\rm tot}$ at the initial stage of the collision and leads to the peak 
of the ratio at about 5 fm/c. At $t > 10$ fm/c, the meson-nucleon collisions dominate. Thus, 
$<\sigma_{12}^\star>$ is close to 20 mb in both calculations, with vacuum and in-medium baryon-baryon cross 
sections. As a consequence, at $t > 10$ fm/c the ratio $N_3/N_{\rm tot}$, practically, does not depend on the 
baryon-baryon cross sections.

Another interesting feature is that the average $\gamma$-factor (c.f. dotted lines in the top left and right 
panels of Fig.~\ref{fig:freq}) drops from initial value of $3.6$ to a rather low value of $1.5$ within 5 fm/c.
This reflects the transition of the fast relative motion of colliding nuclei to a slower thermal motion of 
nearly equilibrated hadronic matter. The Lorentz contraction of the interaction volume (Eqs. (\ref{V12}) and 
(\ref{V12_aver})) plays only a moderate role for the equilibrated hadronic medium. However, at the initial 
nonequilibrium stage, the Lorentz contraction suppresses the ratio $N_3/N_{\rm tot}$ very strongly. Thus, as 
it should be, the very early stage of a relativistic heavy ion collision can be described by the binary 
cascade model quite well.

\begin{figure}

\includegraphics{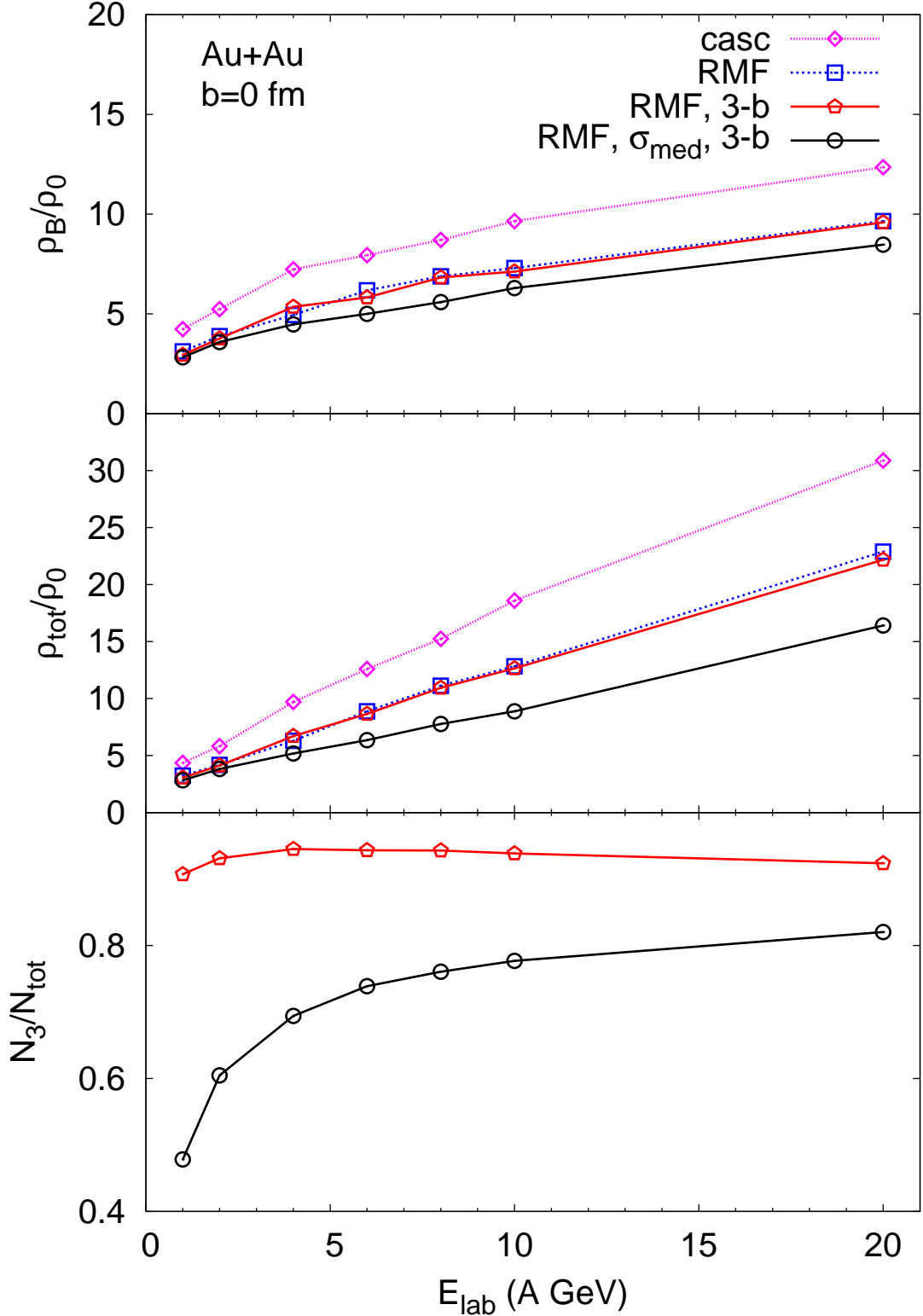}

\vspace*{1cm}

\caption{\label{fig:ratio} (color online) Top, middle and bottom panels, respectively: the maximum central baryon density, 
the maximum central total (i.e. baryon plus meson) density  and the maximum ratio of the three-body collision frequency to 
the total (two- plus three-body) collision frequency reached in the central Au+Au collision vs the beam energy.
Various calculations are shown with the same notations as in Fig.~\ref{fig:freq_tot}.}

\end{figure}

Fig.~\ref{fig:ratio} shows the beam energy dependence of the maximum central baryon and total densities and 
of the maximum ratio $N_3/N_{\rm tot}$ reached in central Au+Au collisions. RMF reduces the maximum baryon 
and total densities quite substantially (c.f. also Fig.~\ref{fig:freq_tot}). The three-body collisions do 
not influence these two observables. 
In calculation with the vacuum baryon-baryon cross sections, the ratio $N_3/N_{\rm tot}$ stays almost constant
$\sim 0.9$ in the beam energy range from 2 to 20 A GeV. Using the in-medium baryon-baryon cross sections leads 
to a dropping $N_3/N_{\rm tot}$ towards smaller beam energies. Indeed, at smaller $E_{\rm lab}$, the 
in-medium reduction of the baryon-baryon cross sections is better visible, due to the smaller number 
of the meson-baryon collisions. 

\subsection{ Comparison with experiment}
 
\begin{figure}

\includegraphics{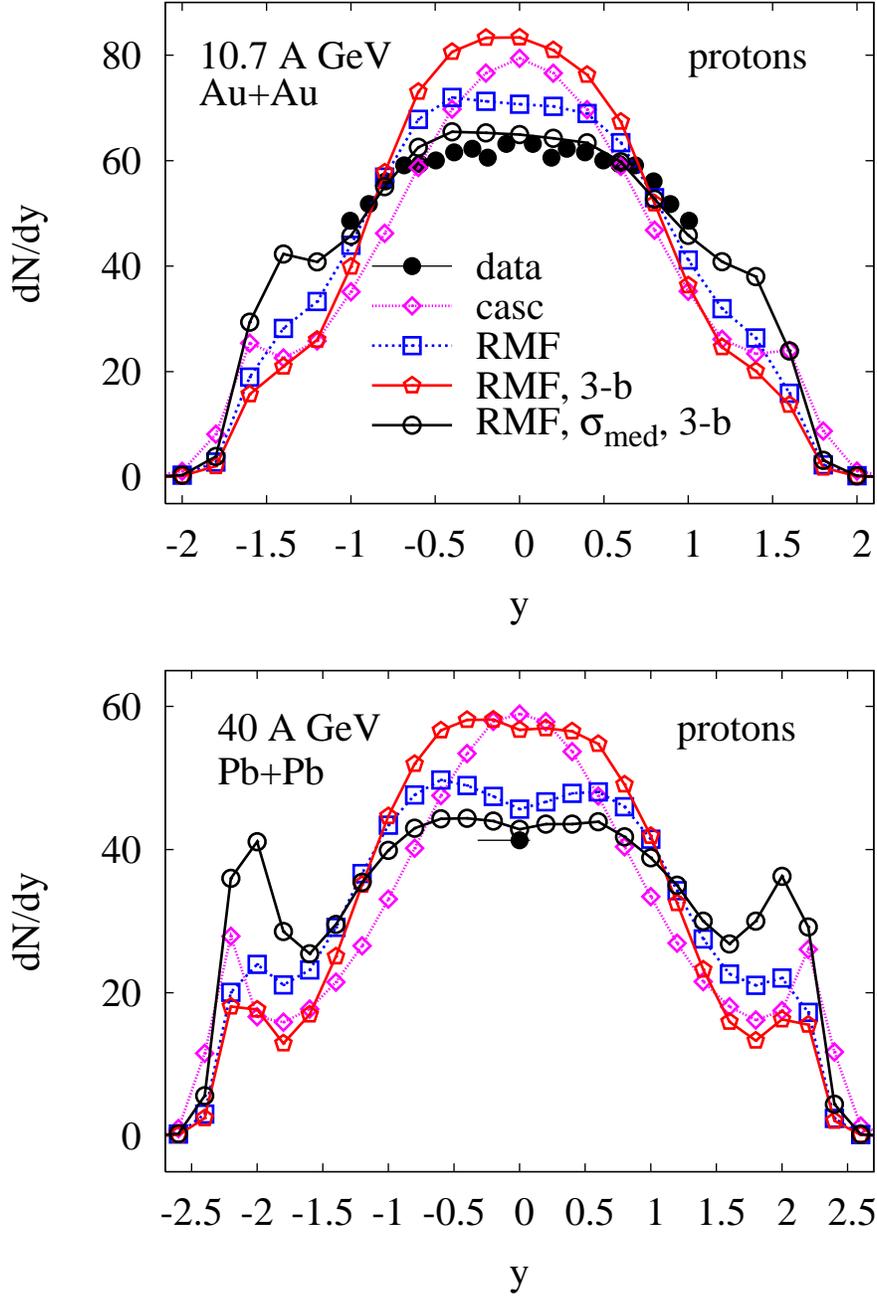}

\caption{\label{fig:dNdy} (color online) Proton rapidity disributions for 
central ($b \leq 3.5$ fm) Au+Au collisions at 10.7 A GeV 
(upper panel), and central ($b \leq 4$ fm) Pb+Pb collisions
at 40 A GeV (lower panel). The experimental data for the Au+Au system
are taken from Ref. \cite{Back02} and correspond to 5\% most central
events. The data for Pb+Pb are from Ref. \cite{Anticic04} (7\% centrality).
Notations are the same as in Fig.~\ref{fig:freq_tot}.}

\end{figure}

First, we address the stopping power of nuclear matter. Fig.~\ref{fig:dNdy} shows the proton rapidity
distributions for the central Au+Au collisions at 10.7 A GeV (upper panel) and for the central Pb+Pb
collisions at 40 A GeV (lower panel). The cascade calculation (i) clearly produces too much stopping.
The same efect has also been observed in earlier GiBUU calculations \cite{WLM05}.

Including the RMF reduces the stopping power. This brings the calculation (ii) into closer agreement with 
the midrapidity proton yields. It is interesting, 
that, at 40 A GeV, the proton rapidity distribution even develops a minimum at $y=0$ indicating an onset of
the transparency. Taking into account three-body collisions (iii) increases the stopping power strongly,
which again results in an overestimation of the midrapidity proton yields. Ultimately, using in-medium
cross sections (iv) reduces the stopping power in a good agreement with the data.

The in-medium reduced baryon-baryon cross sections reduce the stopping power due to less thermalization, 
while the three-body collisions act in the opposite direction leading to more thermalization. Indeed,
a third particle found in the vicinity of the colliding pair interchanges
its energy and momentum with the pair: we simulated this effect by microcanonical sampling of the three-body
phase space (c.f. Eq.(\ref{Prob}). As a result, the relative momentum of the two colliding particles changes 
its direction according to the isotropic distribution. Therefore, the outgoing particles are also produced 
isotropically in the three-body c.m. frame. This has a strong effect on the stopping power, since the 
particle production in hadron-hadron collisions is forward-backward peaked in the c.m. frame of the 
colliding particles at high invariant energies.

\begin{figure}

\includegraphics{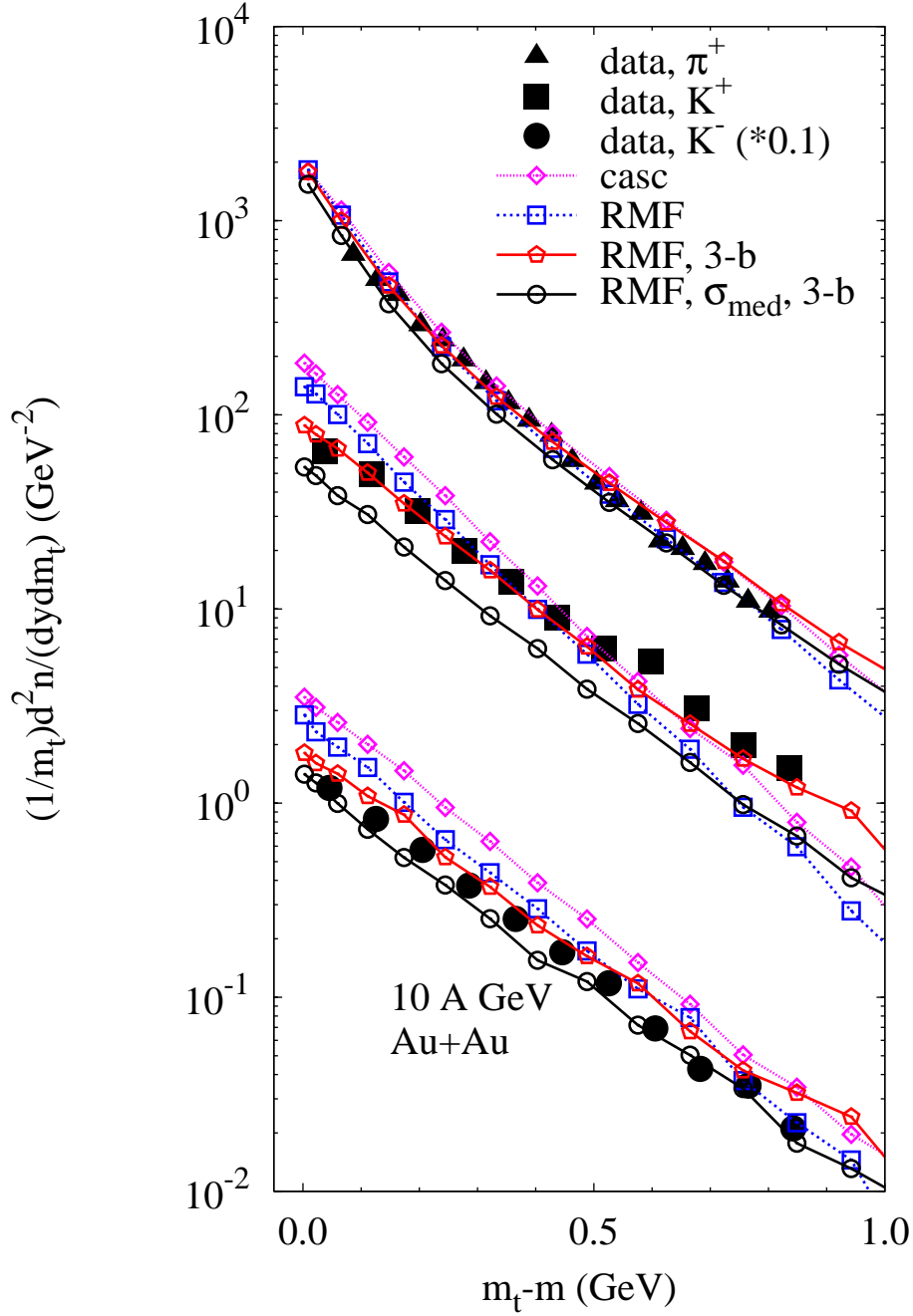}

\caption{\label{fig:dNdmt_10AGeV} (color online) $\pi^+$, K$^+$ and K$^-$ transverse mass spectra at 
midrapidity from central ($b \leq 3.5$ fm) Au+Au collisions at 10.7 A GeV. The rapidity region is 
$|(y-y_{NN})/y_{NN}|<0.125$, where $y_{NN}$ is the c.m. rapidity in the laboratory frame. The K$^-$ 
spectra are multiplied by 0.1. The data are from Refs. \cite{Ahle00_1,Ahle00_2}.
Notations are the same as in Fig.~\ref{fig:freq_tot}.}

\end{figure}

\begin{figure}

\includegraphics{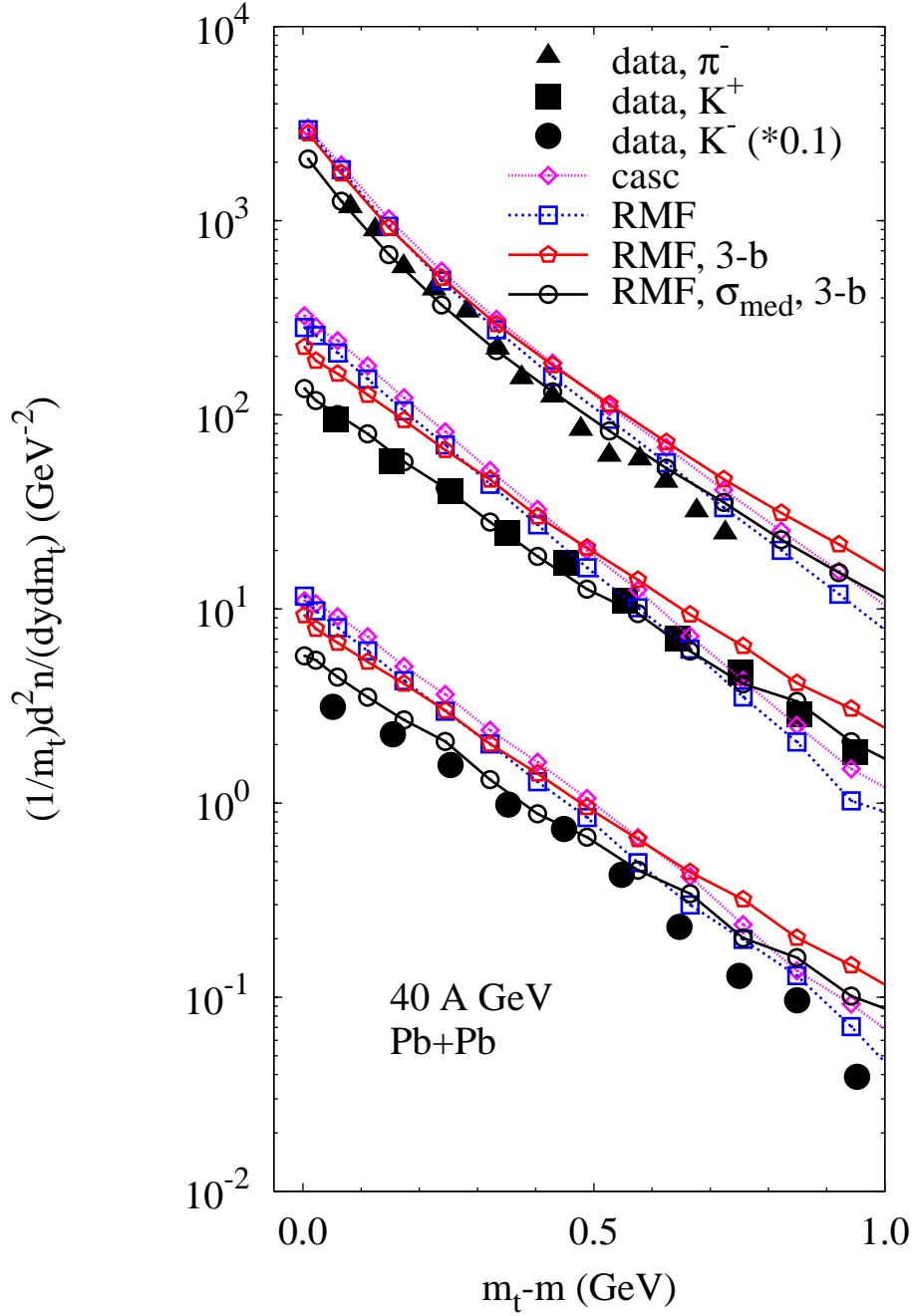}

\caption{\label{fig:dNdmt_40AGeV} (color online) $\pi^-$, K$^+$ and K$^-$ transverse mass spectra at 
midrapidity from central ($b \leq 4$ fm) Pb+Pb collisions at 40 A GeV. The rapidity region is 
$|y-y_{NN}|<0.2$ for $\pi^-$ and $|y-y_{NN}|<0.1$ for K$^\pm$. The K$^-$ spectra are multiplied by 0.1. 
The data are from Ref. \cite{Afan02}.
Notations are the same as in Fig.~\ref{fig:freq_tot}.}

\end{figure}

The effect of three-body collisions can be even better demonstrated by plotting the transverse mass spectra
of the produced mesons.  Figs.~\ref{fig:dNdmt_10AGeV} and \ref{fig:dNdmt_40AGeV} show the $m_\perp$-spectra
of pions, kaons and antikaons at midrapidity produced in central Au+Au collisions at 10.7 A GeV and central 
Pb+Pb collisions at 40 A GeV. The cascade calculation (i) overestimates the meson yields at
midrapidity and also produces too soft $K^+$ $m_\perp$-spectra. Including the RMF into propagation of the
baryons (ii) reduces the yields somewhat, but does not change much the slopes. The three-body collisions
(iii) make the  $m_\perp$-spectra considerably harder. This is again caused by the isotropic emission of the
produced mesons in the c.m. frame of the colliding triple. Applying the in-medium reduced baryon-baryon
cross sections (iv) reduces the meson yields without strong changes of the spectra shapes. 

The last fact is unexpected, since reducing the cross section should also reduce the interaction 
volume, where the third particle is looked for, and, hence, reduce the relative fraction of the
three-body collisions with respect to two-body collisions. However, this is true only for the baryon-baryon
collisions, because we used the in-medium cross sections in this case only.
Therefore, the reduction of the number of the three-body collisions of 
the type baryon-baryon-(baryon or meson) is compensated by the increase of the number of the collisions
of the type meson-baryon-(baryon or meson). As a result, the slopes of the $m_\perp$-spectra did not get 
softer after application of the in-medium baryon-baryon cross sections.

In the case of Au+Au collisions at 10.7 A GeV, the calculation with three-body collisions and vacuum
cross sections (iii) provides the best agreement with the experimental transverse mass spectra of 
$\pi^+$, $K^+$ and $K^-$.  
For the Pb+Pb system at 40 A GeV, the three-body collisions combined with the in-medium cross sections
(iv) produce the best description of the data on $m_\perp$-spectra  of $\pi^-$, $K^+$ and $K^-$.

\begin{figure}

\includegraphics{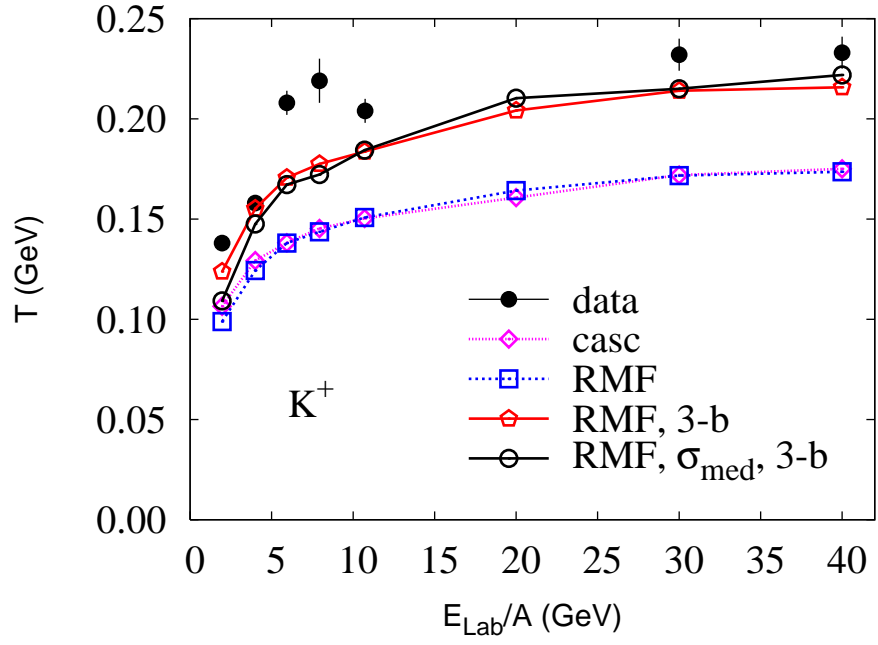}

\caption{\label{fig:temp} (color online) Inverse slope parameter of the $K^+$ transverse mass spectrum at
midrapidity for central collisions of Au+Au and Pb+Pb as function of the beam energy.
Data from Refs. \cite{Ahle00_2,Afan02,Friese04}.
Notations are the same as in Fig.~\ref{fig:freq_tot}.}

\end{figure}

In Fig.~\ref{fig:temp} we present the inverse slope parameter $T$ of the $K^+$ transverse mass spectrum
at midrapidity vs the beam energy. To obtain $T$, following Refs. \cite{Ahle00_2,Afan02,Friese04},
we fitted the midrapidity transverse mass spectrum by an exponential function:
\begin{equation}
   {d^2 n \over m_\perp d m_\perp d y} 
 = a \exp\{-m_\perp/T\}~.                                                    \label{fit}
\end{equation}
Without the three-body collisions, we underpredict the inverse slope parameter
$T$ by about 30\%. Including the three-body collisions leads to the much better agreement with
experiment, except for the points at 5.93 and 7.94 A GeV, where we still underpredict the experimental
inverse slope parameter by about 20\%. 

\begin{figure}

\includegraphics{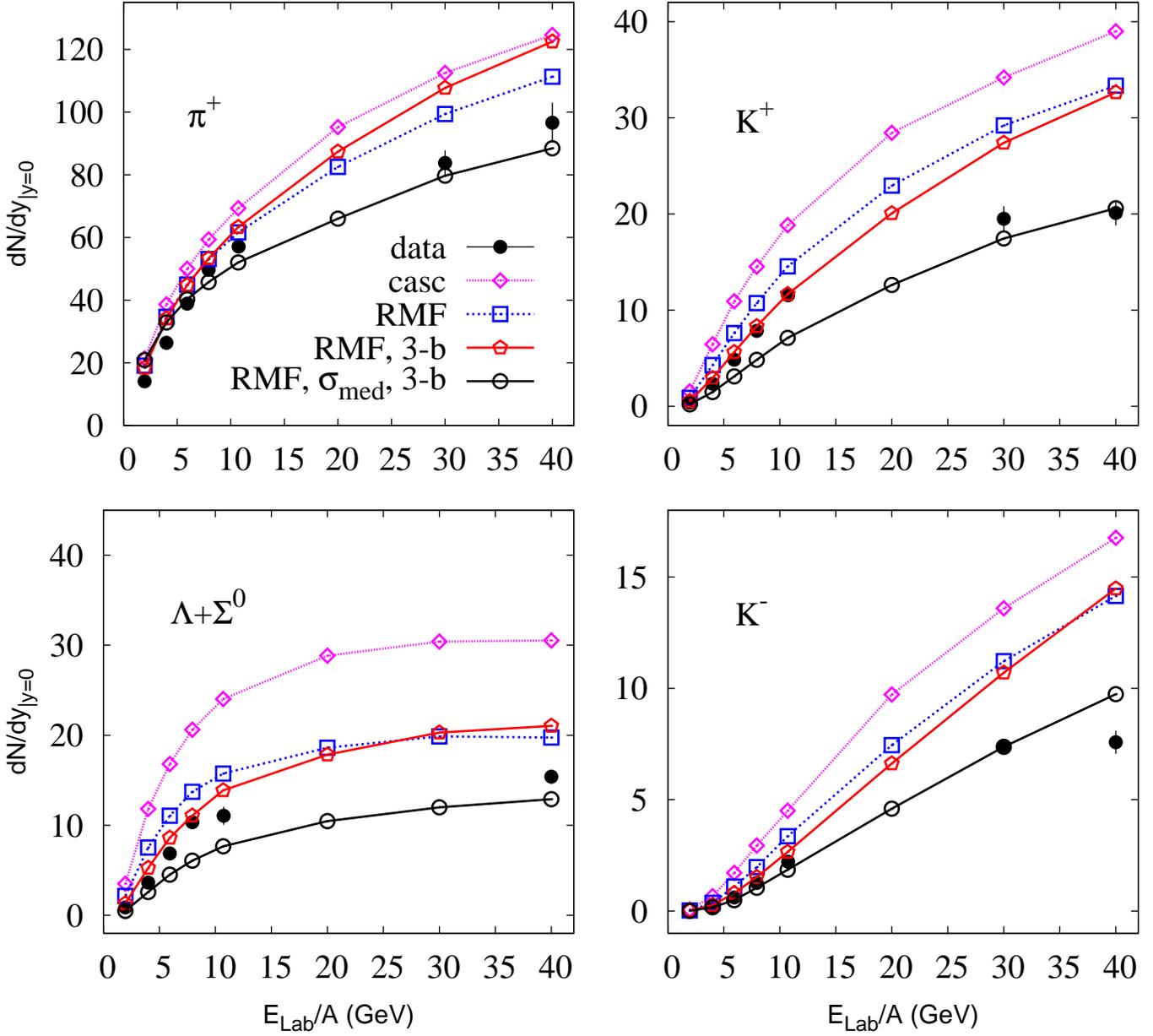}

\caption{\label{fig:dNdy_midr} (color online) The yield of $\pi^+$ (upper left panel), $K^+$ 
(upper right panel), $\Lambda+\Sigma^0$ (lower left panel) and $K^-$ (lower right panel) at 
midrapidity as function of the beam energy for central collisions of Au+Au at $E_{lab} \leq 20$ A GeV 
and Pb+Pb at $E_{lab} = 30$ and 40 A GeV. The data are from 
Refs. \cite{Ahle00_1,Ahle00_2,Afan02,Friese04,Mischke02,Mischke03,Ahmad96,Pink02,Anti99}.
Notations are the same as in Fig.~\ref{fig:freq_tot}.}

\end{figure}

Fig.~\ref{fig:dNdy_midr} shows the midrapidity yields of $\pi^+$, $K^+$, $(\Lambda+\Sigma^0)$ 
and $K^-$ vs the beam energy for central Au+Au collisions at 1.96, 4.00, 5.93, 7.94, 10.7 A GeV
and 20 A GeV and for central Pb+Pb collisions at 30 and 40 A GeV.  The corresponding experimental data on 
pion, kaon, antikaon and hyperon production were taken from Refs. \cite{Ahle00_1,Ahle00_2,Afan02,Friese04,
Mischke02,Mischke03,Ahmad96,Pink02,Anti99}. The calculations for the Au+Au system were done for the impact
parameter range $b \leq 3.5$ fm (5\% of the geometrical cross section, c.f. \cite{Ahle00_1}).
For the Pb+Pb system, we have chosen a slightly larger impact parameter range $b \leq 4$ fm
(7\% of the geometrical cross section, c.f. \cite{Afan02}). 

We observe that the pure cascade calculation (i) overestimates the meson and hyperon production. 
Using the RMF (ii) reduces the midrapidity meson yields by about 15\%. The midrapidity hyperon yield is 
reduced stronger, by about 30\%. This reflects the behaviour of the proton midrapidity yield shown in 
Fig.~\ref{fig:dNdy}, since the mean field acts on the hyperons too. The introduction of the three-body 
collisions (iii) influences the midrapidity yields of the produced particles rather weakly. Finally, the 
in-medium cross sections (iv) reduce the particle production quite strongly: mesons --- by about 30\%, 
and hyperons --- by about 50\%. As a result the calculation (iv) turns out to be in a good agreement with 
the data on pion and $K^-$ production, while it underestimates the $K^+$ and hyperon yields below 40 A GeV. 

\begin{figure}

\includegraphics{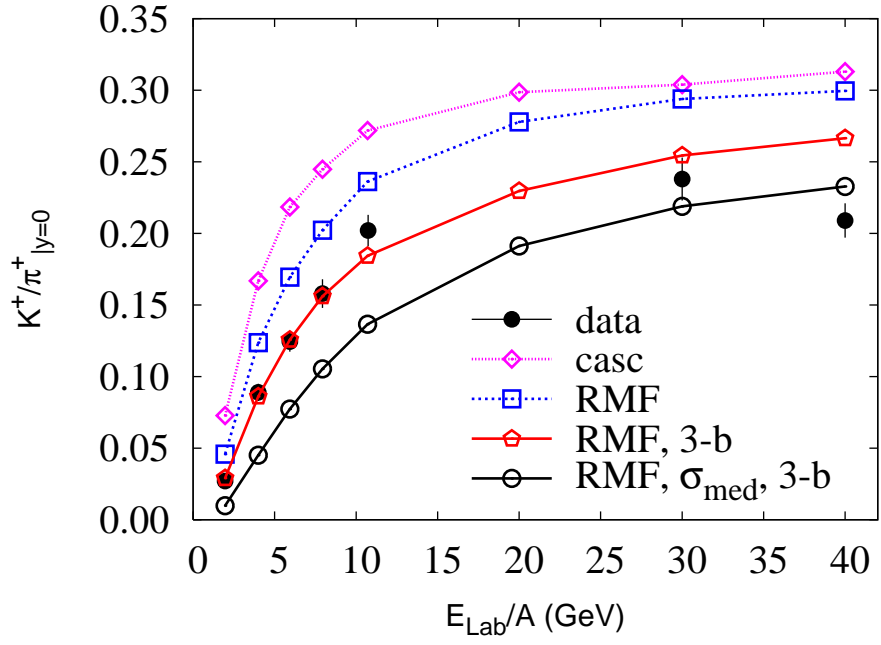}

\caption{\label{fig:kp2pip_ratio} (color online) The ratio of the midrapidity yields $K^+/\pi^+$ for
central Au+Au and Pb+Pb collisions. The data are from \cite{Ahle00_1,Afan02,Friese04}.
Notations are the same as in Fig.~\ref{fig:freq_tot}.}

\end{figure}

Fig.~\ref{fig:kp2pip_ratio} shows the $K^+/\pi^+$ ratio at midrapidity vs the beam energy. It is
interesting, that the three-body collisions reduce the ratio quite strongly. This is due to
combination of the two small effects visible in Fig.~\ref{fig:dNdy_midr}: increase of
the pion yield and decrease of the kaon yield by the three-body collisions. The calculation
in the RMF mode with vacuum cross sections and three-body collisions (iii) is in the best
agreement with the experimental data below 40 A GeV. However, we fail to describe the reduction
of the $K^+/\pi^+$ ratio above 30 A GeV.

\section{Summary and discussion}

We studied the influence of several many-body effects on particle production
in heavy ion collisions at 2-40 A GeV. The calculations were done in the framework of the
GiBUU model \cite{GiBUU}. 

First, we implemented the relativistic mean field of the nonlinear Walecka model NL2 \cite{LCMW92}
in GiBUU. The RMF reduces the stopping power of colliding nuclei and the meson production. 
In a calculation with RMF, a part of the kinetic c.m. energy of the colliding
nuclei transformes into the build-up of the strongly repulsive time component of the $\omega$-field.
This leads to less compression and less entropy production by particle-particle collisions.
As a consequence, the nuclear matter becomes more transparent when the mean field is taken
into account (c.f. Fig.~\ref{fig:dNdy}). This is an interesting result, since in microscopic 
transport models the mean field has been usually not taken into account (or switched-off) 
at high energies (c.f. Refs. \cite{Weber03,Brat04,WLM05}).

Second, we implemented the three-body collisions in the model by adopting the geometrical method 
of Refs. \cite{Mrow85,BRV92,BM93,BGM94} modified to account for the relativistic effects (see Sect.IIC). 
The three-body collisions increase the inverse slope parameter of the $K^+$ transverse mass spectra to 
a quite good agreement (within 10\%) with experimental data, except for the points at 5.93 and 7.94 A GeV.
The additonal transverse momentum is generated due to isotropic emission of the produced
particles in the three-body c.m. frame.  

The RMF model serves also as a natural base for the construction of in-medium modified baryon-baryon
cross sections. Here, we continued a study started in Ref. \cite{WLM05}. Assuming the matrix
element of the meson production in a baryon-baryon collision to be not modified by the nuclear medium,
we considered the in-medium modifications of phase space and flux factors. We also
took into account the in-medium normalization of the Dirac plane wave bispinor. The last effect
has led to a strong in-medium reduction of the cross sections. However, the way we extract
the in-medium cross sections is very approximate, since it neglects in-medium modifications
of the matrix element.  

We have also studied the $\pi^+$, $K^+$, $K^-$ and $(\Lambda+\Sigma^0)$-hyperon yields at midrapidity
at various beam energies (Fig.~\ref{fig:dNdy_midr}). Given the ambiguity in the in-medium cross sections,
the data are reasonably well described by calculations with the three-body collisions. The same
is valid for the $K^+/\pi^+$ ratio at midrapidity plotted vs the beam energy (Fig.~\ref{fig:kp2pip_ratio}).
However, we do not reproduce the decrease of the $K^+/\pi^+$ ratio above 30 A GeV.

The problems of the microscopic transport models to describe the $K^+/\pi^+$ ratio have usually been 
ascribed to the excessive pion yield, while the strangeness production was well reproduced overall
\cite{Weber03,Brat04,WLM05}. This conclusion was based on calculations within the cascade mode
using vacuum cross sections. In the present work, by using the in-medium reduced cross sections,
we have well reproduced the $\pi^+$ midrapidity yields, while the $K^+$ midrapidity yields are
now underpredicted.  Since pions represent the major contribution to particle production, we believe
that the in-medium cross sections, nevertheless, provide a reasonable base for description of the
energy transfer to the inelastic channels. The $K^+$ yield, thus, can be enhanced by more detailed
elaboration on the strangeness production channels: $BB \to BYK$ channel below FRITIOF threshold,
larger phenomenological strangeness suppression factor $\gamma_s$ at small $\sqrt{s}$ (c.f. Eq.(\ref{strsuppr})).
Moreover, time dependent prehadron cross sections of Ref. \cite{GM07} could also influence the results
on strangeness production. These topics deserve, in our opinion, future studies within the GiBUU or similar 
transport approaches.

In spite of the fundamental problems with the kinetic theory at high densities, we believe that
our approach combining the baryon propagation in RMF with the in-medium reduced cross sections
and three-body collisions provides a realistic description of the HIC dynamics in 10 A GeV domain.
It could also serve to model the pre- and after-quark-gluon phase formation stages of a HIC
at higher energies.   

\begin{acknowledgments}
We gratefully acknowledge support by the Frankfurt Center for Scientific
Computing. We also acknowledge helpful discussions with Dr. T. Gaitanos on the relativistic mean
field model. One of us (A.B.L.) is grateful to Prof. I.N. Mishustin and Prof. P. Senger for stimulating 
discussions and useful comments.
\end{acknowledgments}

\appendix

\section{Elastic three-body matrix element}

It is straightforward to write down the three-body elastic collision integral which corresponds to the 
procedure discussed in Sect. IIC:
\begin{eqnarray}
  I_{\rm coll, 3b}^{\rm elastic}[f_1^\star] &=& \int\, \frac{ g d^3 p_2^\star }{ (2\pi)^3 } %
  \sigma_{12}^\star v_{12}^\star V_{12} %
  \int\, \frac{ g d^3 p_3^\star (\gamma_3)_{\rm cm12} }{ (2\pi)^3 \gamma_3 }\,  %
  \Phi_3^{-1}(\sqrt{s_{123}^\star}; m_{1^\prime}^\star, m_{2^\prime}^\star, m_{3^\prime}^\star)  \nonumber \\
&\times& \int\, d\Phi_3(p_1^\star+p_2^\star+p_3^\star;\,%
                        p_{1^\prime}^\star,\, p_{2^\prime}^\star,\, p_{3^\prime}^\star)\,%
        \sigma_{1^\prime2^\prime}^{\star\,-1} % 
        \int\, d\sigma_{ 1^\prime 2^\prime \to 1^{\prime\prime} 2^{\prime\prime} }^\star  \nonumber \\
&\times& (  f_{1^{\prime\prime}}^\star f_{2^{\prime\prime}}^\star  f_{3^\prime}^\star % 
            \bar{f}_1^\star \bar{f}_2^\star \bar{f}_3^\star  %
          - f_1^\star f_2^\star f_3^\star %
            \bar{f}_{1^{\prime\prime}}^\star \bar{f}_{2^{\prime\prime}}^\star \bar{f}_{3^\prime}^\star )~.
                                                                     \label{I3el_exp0}
\end{eqnarray}
Here $\sigma_{12}^\star$ and $v_{12}^\star$ are, respectively, the total (in-medium) interaction cross 
section and the relative velocity (Eq.(\ref{v12Star})) of 1 and 2;
\begin{equation}
   V_{12} = \frac{4}{3} \pi (R_{\rm max})^3 \gamma_{12}^{-1}                      \label{V12}
\end{equation}
is the interaction volume, where $R_{\rm max}$ and $\gamma_{12}$ are given by Eqs. (\ref{Rmax}) and 
(\ref{gamma12}), respectively; $\gamma_3 = p_3^{*0}/m_3^*$;
$\Phi_3$ and $d\Phi_3$ are the three-body phase space volume (Eq.(\ref{Phi})) and the three-body phase
space volume element (Eq.(\ref{dPhi})), respectively;
$s_{123}^\star \equiv (p_1^\star+p_2^\star+p_3^\star)^2$;
$d\sigma_{ 1^\prime 2^\prime \to 1^{\prime\prime} 2^{\prime\prime} }^\star$ is the elastic 
differential scattering cross section (Eq.(\ref{dSigmaStar})). The quantity 
\begin{equation}
   \int\, \frac{ g d^3 p_3^\star (\gamma_3)_{\rm cm12} }{ (2\pi)^3 \gamma_3 } f_3^\star     \label{dens_cm12}
\end{equation}
is the density of particles in the c.m. frame of the particles $1$ and $2$.

Notice, that Eq.(\ref{I3el_exp0}) takes into account the two transitions: 
$1 2 3 \to 1^\prime 2^\prime 3^\prime$ by the microcanonical sampling the kinetic momenta of
$1^\prime, 2^\prime$ and $3^\prime$ according to Eq.(\ref{Prob}), and the elastic scattering
$1^\prime2^\prime \to 1^{\prime\prime}2^{\prime\prime}$ afterwards.
The divisions by $\Phi_3$ and by $\sigma_{1^\prime2^\prime}^\star$ are done in Eq.(\ref{I3el_exp0}) 
in order to normalize the total transition probabilities to the unity and to the branching
ratio of the elastic channel $\sigma_{1^\prime 2^\prime}^{{\rm elastic}\,\star}/\sigma_{12}^\star$,
respectively. Here $\sigma_{1^\prime 2^\prime}^{{\rm elastic}\,\star}=%
\int\, d\sigma_{ 1^\prime 2^\prime \to 1^{\prime\prime} 2^{\prime\prime} }^\star$.

Subsututing the explicit expressions for $d\Phi_3(p_1^\star + p_2^\star + p_3^\star;\,% 
p_{1^\prime}^\star,\, p_{2^\prime}^\star,\,  p_{3^\prime}^\star)$
and for $d\sigma_{ 1^\prime 2^\prime \to 1^{\prime\prime} 2^{\prime\prime} }^\star$
obtained, respectively,  from Eqs. (\ref{dPhi}) and (\ref{dSigmaStar})   
into Eq.(\ref{I3el_exp0}) and changing the order of integrations, one can transform Eq.(\ref{I3el_exp0}) 
to the form of Eq.(\ref{I3el}):
\begin{eqnarray}
  I_{\rm coll, 3b}^{\rm elastic}[f_1^\star] &=& \int\, \frac{ g d^3 p_2^\star }{ (2\pi)^3 } %
  \sigma_{12}^\star v_{12}^\star V_{12} %
  \int\, \frac{ g d^3 p_3^\star (\gamma_3)_{\rm cm12} }{ (2\pi)^3 \gamma_3 }\,  %
   \Phi_3^{-1}(\sqrt{s_{123}^\star}; m_{1^\prime}^\star, m_{2^\prime}^\star, m_{3^\prime}^\star)  \nonumber \\
&\times& \int\, \frac{ d^3 p_{1^\prime}^\star }{ (2\pi)^3 2p_{1^\prime}^{\star0} } %
         \int\, \frac{ d^3 p_{2^\prime}^\star }{ (2\pi)^3 2p_{2^\prime}^{\star0} } %
         \int\, \frac{ d^3 p_{3^\prime}^\star }{ (2\pi)^3 2p_{3^\prime}^{\star0} } %
         \delta^{(4)}(p_1^\star+p_2^\star+p_3^\star-p_{1^\prime}^\star-p_{2^\prime}^\star-p_{3^\prime}^\star) %
                                                                                                  \nonumber \\      
&\times&  \sigma_{1^\prime2^\prime}^{\star\,-1}\,\sigma_{1^\prime 2^\prime}^{{\rm elastic}\,\star}\,%
          (  f_{1^\prime}^\star f_{2^\prime}^\star  f_{3^\prime}^\star \bar{f}_1^\star \bar{f}_2^\star \bar{f}_3^\star  %
          - f_1^\star f_2^\star f_3^\star \bar{f}_{1^\prime}^\star \bar{f}_{2^\prime}^\star \bar{f}_{3^\prime}^\star )~.
                                                                     \label{I3el_exp}
\end{eqnarray}
In derivation of Eq.(\ref{I3el_exp}), we used the relations
\begin{eqnarray}
   \sigma_{1^\prime2^\prime}^\star &=& \sigma_{1^{\prime\prime}2^{\prime\prime}}^\star~,   \label{relation1}\\
   I_{1^\prime2^\prime}^\star &=& I_{1^{\prime\prime}2^{\prime\prime}}^\star~,             \label{relation2}   
\end{eqnarray}
valid for the elastic scattering $1^\prime2^\prime \to 1^{\prime\prime}2^{\prime\prime}$ and
the detailed balance formula
\begin{equation}
   \overline{|M|^2_{1^\prime2^\prime \to 1^{\prime\prime}2^{\prime\prime}}} = 
   \overline{|M|^2_{1^{\prime\prime}2^{\prime\prime} \to 1^\prime2^\prime}}~.              \label{detbal1}
\end{equation}
Finally, we have also changed the notations as $1^{\prime\prime} \to 1^\prime$ and
$2^{\prime\prime} \to 2^\prime$.
By putting all the $\gamma$-factors equal to unity and also neglecting the Pauli blocking of the
final states, the loss term of Eq.(\ref{I3el_exp}) is reduced to the loss term of Eq.(4.17) from 
Ref. \cite{BGM94}. The comparison of the gain term of our Eq.(\ref{I3el_exp}) to the gain term
of Eq.(4.17) from Ref. \cite{BGM94} is not straightforward, since we used an assumption of
the microcanonical sampling of the particles $1^\prime$, $2^\prime$ and $3^\prime$ momenta,
which has not been used in Ref. \cite{BGM94}.  

By comparison Eqs. (\ref{I3el_exp}) and (\ref{I3el}) we get the following expression for the invariant 
matrix element of the elastic three-body collisions:
\begin{eqnarray}
   \overline{|M|^2_{123 \to 1^\prime2^\prime3^\prime}} &=& 
   \frac{ 3!\,2!\,\sigma_{12}^\star\,I_{12}^\star\,\frac{4}{3} \pi (R_{\rm max})^3 (\gamma_3)_{\rm cm12} } %
        { m_1^\star m_2^\star m_{1^\prime}^\star m_{2^\prime}^\star m_{3^\prime}^\star (2g)^3 (2\pi)^4 %
          \gamma_{12} } %
   \Phi_3^{-1}(\sqrt{s_{123}^\star}; m_{1^\prime}^\star, m_{2^\prime}^\star, m_{3^\prime}^\star)
   \frac{ \sigma_{1^\prime 2^\prime}^{{\rm elastic}\,\star} }%
        { \sigma_{1^\prime2^\prime}^\star }~.
                                                                                          \label{M2}
\end{eqnarray} 
The ratio of the $\gamma$-factors in the c.m. frame of $1$ and $2$ can be rewritten in the explicitly 
covariant form as
\begin{equation}
  \frac{(\gamma_3)_{\rm cm12}}{\gamma_{12}}  %
  = \frac{ 2 m_1^\star\, m_2^\star\, p_3^\star\, ( p_1^\star + p_2^\star ) } %
         { m_3^\star\, ( m_2^\star\, p_1^\star + m_1^\star\, p_2^\star ) ( p_1^\star + p_2^\star ) }~.   
                                                                                          \label{gamRatio}
\end{equation}

\clearpage

\thispagestyle{empty}

\clearpage

\thispagestyle{empty}

\begin{figure}

\includegraphics{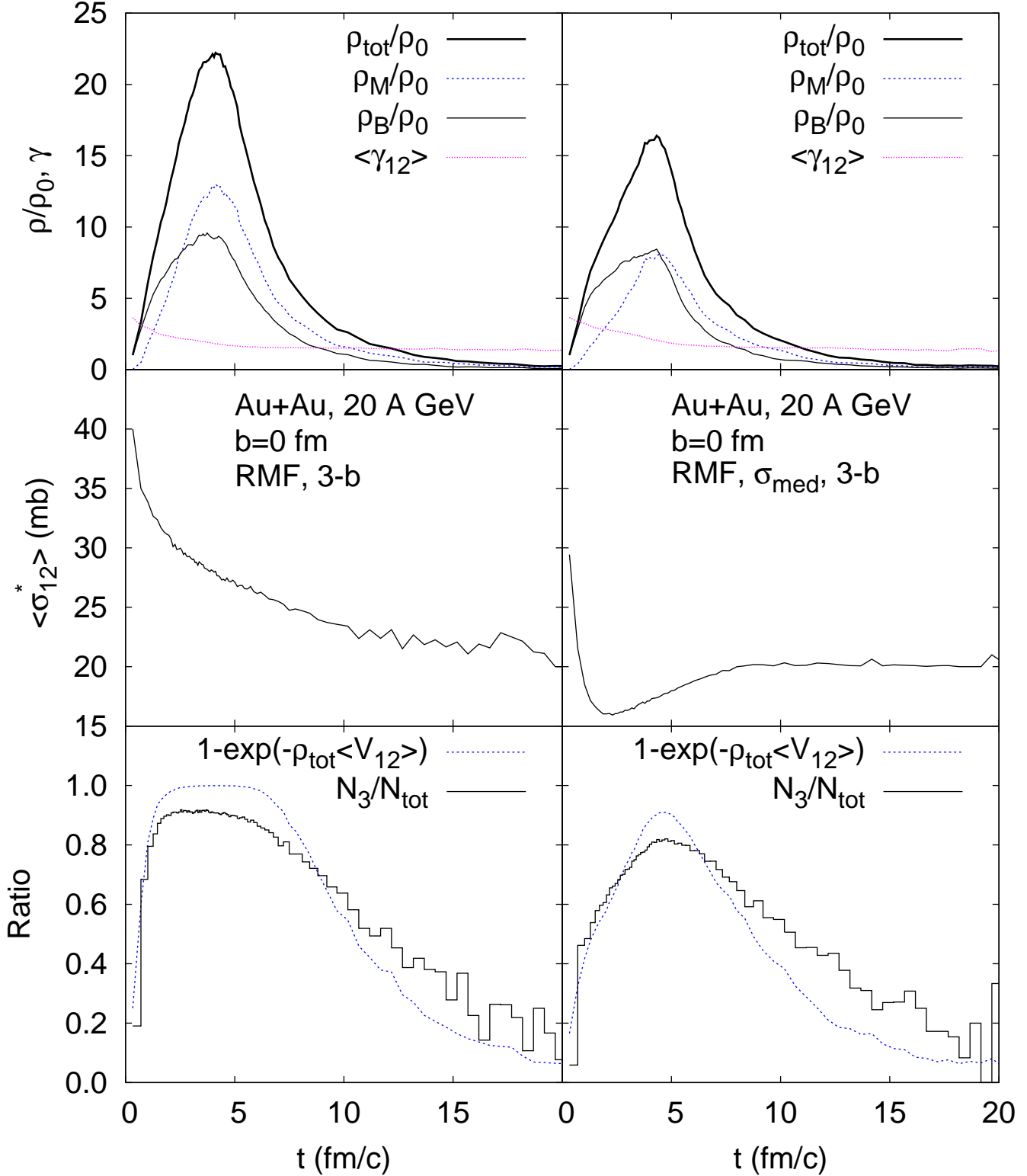}

\vspace*{1cm}

\caption{\label{fig:freq} (color online) Top panels: time dependence of the total (baryon plus meson) density
--- thick solid lines, meson density --- dashed lines, baryon density --- thin solid lines and of
the $\gamma$-factor (Eq.(\ref{gamma12})) averaged over colliding pairs --- dotted lines.
Middle panels: cross section used in calculation of the radius $R_{\rm max}$ (Eq.(\ref{Rmax}))
averaged over colliding pairs. Bottom panels: the ratio of the three-body and
the total (two- plus three-body) collision freequencies --- solid histograms and the estimate of
Eq.(\ref{estimate}) --- dashed lines. The calculations are done with RMF including three-body collisions.
Left column: with vacuum cross sections. Right column: with in-medium cross sections.
The colliding system is Au+Au at 20 A GeV, b=0 fm.}

\end{figure}

\clearpage

\thispagestyle{empty}

\clearpage

\thispagestyle{empty}

\clearpage

\thispagestyle{empty}

\clearpage

\thispagestyle{empty}

\clearpage

\thispagestyle{empty}

\clearpage

\thispagestyle{empty}

\clearpage

\thispagestyle{empty}

\begin{thebibliography}{99}
\bibitem{com1} We would like to remind that we are using the notations of
Ref. \cite{LKR97} for the RMF parameters. In order to go back to the
notations of Ref. \cite{LCMW92}, one needs to do the following replacements
in the Lagrangian density (\ref{Lagr}): $g_\sigma \to g_s$, $g_\omega \to g_v$,
$m_\sigma \to m_s$, $m_\omega \to m_v$, $g_2 \to -B$, $\sigma \to -\sigma$
and $g_3 \to C$.
\bibitem{com2} Except for the case of $3\to2$ collisions, 
where the matrix element can be obtained by using the detailed balance relations from the 
cross section of the inverse process.
\end{thebibliography}
\end{document}